\documentclass[11pt]{article}
\usepackage{graphics,epsfig}
\usepackage[latin1]{inputenc}
\usepackage[english,activeacute]{babel}
\usepackage{epstopdf}
\usepackage{graphicx}
\usepackage{latexsym}
\usepackage{amsmath, amsthm, amsfonts, amssymb}
\usepackage{verbatim}

\begin{document}

\title{On the classical limit of quantum mechanics, fundamental graininess and chaos: compatibility of chaos with the correspondence principle}
\author{\textsc{Ignacio Gomez}$^{1}$ and \textsc{Mario Castagnino}$^{2}$}
\maketitle

\begin{abstract}
The aim of this paper is to review the classical limit of Quantum Mechanics
and to precise the well known threat of chaos (and fundamental graininess)
to the correspondence principle. We will introduce a formalism for this
classical limit that allows us to find the surfaces defined by the constants of
the motion in phase space. Then in the integrable case we will find the
classical trajectories, and in the non-integrable one the fact that regular
initial cells become ``amoeboid-like". This deformations and their
consequences can be considered as a threat to the correspondence principle unless we take into account the characteristic timescales of quantum chaos.
Essentially we present an analysis of the problem similar to the one of
Omn\`{e}s \cite{Omnes,Omnes1}, but with a simpler mathematical structure.
\end{abstract}

\newtheorem{theo}{Theorem}[section] \newtheorem{definition}[theo]{Definition}
\newtheorem{lem}[theo]{Lemma} \newtheorem{prop}[theo]{Proposition} %
\newtheorem{coro}[theo]{Corollary} \newtheorem{exam}[theo]{Example} %
\newtheorem{rema}[theo]{Remark} \newtheorem{example}[theo]{Example} %
\newtheorem{principle}[theo]{Principle} \newtheorem{axiom}[theo]{Axiom}

%\numberwithin{equation}{subsection}

\begin{center}
{\small 1- Instituto de Física de Rosario (IFIR-CONICET), Rosario, Argentina\\[0pt]
2- Instituto de Física de Rosario (IFIR-CONICET) and \\[0pt]
Instituto de Astronomía y Física del Espacio, \\[0pt]
Casilla de Correos 67, Sucursal 28, 1428 Buenos Aires, Argentina.\\[0pt]}
\end{center}

\vspace{1cm}

\bigskip \noindent

{\small \noindent
\centerline{\emph{Key words:
weak-limit-Local CSCO-classical limit-van Hove observables-graininess}}

\section{Introduction}

It seems that Einstein was the first one to realize that chaos was a threat
to quantum mechanics \cite{Ikeda} in a paper that was ignored by forty years
\cite{Gutzwiller}. A panoramic view of the this incompatibility of the
classical chaos and quantum concepts (up to 1994) can be found in \cite{Ikeda} and a recent review in \cite{Landsman}.
Our first contribution to the subject was the introduction of a theory of the classical limit for
closed quantum systems with Hamiltonian with continuous spectrum based in
destructive interference (that we have called the ``Self Induced Decoherence"
-SID- and where we have used the Riemann-Lebesgue theorem \cite{SID}) and
later we found a class of quantum chaotic systems (that may not contain all
cases but certainly it contains the relevant ones) with chaotic classical
limit \cite{CSF} \cite{NACHOSKY MARIO}. With this idea in mind we study quantum chaos in papers
\cite{CSF, NACHOSKY MARIO, QCh} and extended the notions of non-integrable, ergodic and mixing
quantum systems in paper \cite{CL}. These works were inspired in the
landmark paper of Bellot and Earman \cite{BE}. The aim of this remarkable
paper is precisely to show ``how chaos puts some pressure on the
correspondence principle (CP)" and the author says that there is not a
``quick and convincing argument for the conclusion that the CP fails".
Another important source of inspiration for us was the two books of Roland
Omn\`{e}s \cite{Omnes}\ and \cite{Omnes1}, precisely the characterization of
quantum chaos as the evolution of a square cell to a distorted ``amoeboid"
cell (see figure 6.B). In this paper we will essentially follow this idea,
with simpler mathematical methods, and we will try to precise the origin of
the elongated, distorted and final amoeboid cells which, in fact, we
consider the main threat to the CP.
It should be noted that the standard approach of the graininess has already been pointed out both in classical discretized
systems and in quantum mechanics by looking at the Kolmogorov-Sinai entropy
and its quantum variants \cite{crifal1, crifal2, falman, bencap1, bencap2}. In these cases, there is no threat to the correspondence principle, but only the
emergence of a typical time-scale over (logarithmic in $\hbar^{-1}$) that signals the non-
commutativity of the limit $t\rightarrow\infty$ and $\hbar\rightarrow0$. This fact is not really taken as
a threat to the correspondence principle. However, we will see that the ameboid-like behavior involves a ``coarse-grained distribution function", i.e. a point-test-distribution function averaged on rectangular rigid boxes of the phase space %and performed after the time limit $t\rightarrow\infty$
(see section 5). %and therefore it is not just another way there emerges the non-commutativity of the two limits of above.
%In our approach the distribution function is simply the classical distribution $\rho_{\ast}(\phi)$ which the system converges in the phase space in the limit $t\rightarrow\infty$.
This coarse-grain is used to get rid the complicated structure of the phase space which becomes more and more ``scarred" as the relaxation proceeds (see \cite{casati chirikov}, pag. 7). Moreover, for the case of a two dimensional phase space we will show a connection between the characteristic timescales of the quantum chaos and an adimensional parameter $\Omega$ which measures the degree of the deformation of the cells as the system evolves.
%and where the volume of each cell is coarse-grained on a rectangular rigid box $\Delta_{\Theta}\Delta_{J}$.

The paper is organized as follows. Section 2: We introduce the mathematical structures we will use.
In the next sections we will see that the classical limit can be obtained using three weapons: decoherence, Wigner transformation and the limit $\frac{\hbar}{S}\rightarrow 0$. Section 3: We review the decoherence alla SID for non-integrable quantum
systems. Section 4: We obtain the classical statistical limit, using Wigner
transformation and the limit $\frac{\hbar }{S}\rightarrow 0$, and the
classical surfaces defined by the constant of the motion in phase space. Section 5: Deals the graininess of quantum mechanics. We find the classical trajectories for the integrable system and
estimate the threat to the CP, in the non-integrable case. We
show that, up to this point the threat to the CP can be suppressed if we take into account the characteristic timescales of quantum chaos. Moreover, we analyze how the fundamental graininess improves the statistical classical limit of section 4.
Section 6: We present our conclusions.

\section{Mathematical background}

In this section we will review, following ref. \cite{CSF} \cite{NACHOSKY MARIO}, the main
mathematical concepts we will use in these papers.

\subsection{Weak limit}

Our presentation is based on the algebraic formalism of quantum mechanics (%
\cite{Emch}, \cite{Haag}). Let us consider an algebra $\mathcal{A}$ of
operators, whose self-adjoint elements $O=O^{\dagger }$ are the observables
belonging to the space $\mathcal{O}$. The states $\rho $ are linear
functionals belonging to the dual space $\mathcal{O}^{\prime }$, but they
must satisfy the usual conditions: self-adjointness, positivity and
normalization and therefore the state $\rho $ belongs to a convex $\mathcal{S%
}$. If $\mathcal{A}$ is a C*-algebra, it can be represented by a Hilbert
space (GNS theorem see \cite{Haag}). If $\mathcal{A}$ is a nuclear algebra,
it can be represented by a rigged Hilbert space, as proved by a
generalization of the GNS theorem (\cite{Iguri-99}, \cite{Iguri-08}). In
this case, the van Hove states with a singular diagonal can be properly
defined (see \cite{vH}; for a rigorous presentation of the formalism, see
also \cite{Antoniou}).

If we write the action of the functional $\rho $ on the space $\mathcal{O}$
as $(\rho |O)$, then we can say that:

\begin{itemize}
\item The evolution $U_{t}\rho =\rho (t)$ has a \textit{Weak-limit} if, for
any $O\in \mathcal{O}$ and any $\rho \in \mathcal{S}$, there is a unique $%
\rho _{\ast }\in \mathcal{S}$ such that
\begin{equation}
\lim_{t\rightarrow \infty }(\rho (t)|O)=(\rho _{\ast }|O),\text{ \ \ \ \ }%
\forall \text{ }O\in \mathcal{O}  \label{2.1}
\end{equation}%
We will symbolize this limit as
\begin{equation}
W-\lim_{t\rightarrow \infty }\rho (t)=\rho _{\ast }  \label{2.2}
\end{equation}

\item A particular useful weak limit can be obtained using the
Riemann-Lebesgue theorem. The idea of destructive interference is embodied
in this theorem, according to which, if $f(\nu )\in \mathbb{L}_{1}$, then
\begin{equation}
\lim_{t\rightarrow \infty }\int_{a}^{b}f(\nu )\;e^{-i\nu t}\;d\nu =0
\label{2.5}
\end{equation}

If we can express the action of a functional $\rho (t)\in \mathcal{S}$ on
the operator $O\in \mathcal{O}$ as
\begin{equation}
(\rho (t)|O)=\int_{a}^{b}\left[ A\,\delta (\nu )+f(\nu )\right] \;e^{-i\nu
t}\;d\nu  \label{2.6}
\end{equation}%
with $f(\nu )\in \mathbb{L}_{1}$, then
\begin{equation}
\lim_{t\rightarrow \infty }(\rho (t)|O)=\lim_{t\rightarrow \infty
}\int_{a}^{b}\left[ A\,\delta (\nu )+f(\nu )\right] \;e^{-i\nu t}\;d\nu
=A=(\rho _{\ast }|O),\text{ \ \ }\forall \text{ }O\in \mathcal{O}
\label{2.7}
\end{equation}%
We will call this result \textquotedblleft \textit{Weak Riemann-Lebesgue
limit}\textquotedblright .
\end{itemize}

\subsection{Generalized Projections.}

As it is well known, in order to describe an irreversible process in terms
of an unitary evolution it is necessary to break the underlying unitary
evolution. \ The usual tool to do this is to introduce a coarse graining,
that restricts the information of the system. But generically any
information restriction can be obtained using a projection, which retains
the \textquotedblleft relevant\textquotedblright\ information and discards
the \textquotedblleft irrelevant\textquotedblright\ one of the considered
system.

In fact, in its traditional form, the action of a projection is to eliminate
some components of the state vector corresponding to the finest description
(see \cite{Mackey}) to obtain a coarse grained one. If this idea is
generalized, any restriction of information can be conceived as the result
of a convenient projection. In fact, we can define a projector $\Pi $
belonging to the space $\mathcal{O}\otimes \mathcal{O}^{\prime }$ such that

\begin{equation}
\Pi \circeq \sum_{j}|O_{j})(\,\rho _{j}|  \label{2.10}
\end{equation}%
where ($\rho _{j}|\in \mathcal{O}^{\prime }$ satisfies $(\rho
_{j}|O_{k})=\delta _{jk}$ where $|O_{k})\epsilon \mathcal{O}$ \footnote{%
In fact, $\Pi $ is a projector since
\begin{equation*}
\Pi ^{2}=\sum_{jk}|O_{j})(\,\rho _{j}|O_{k})(\,\rho
_{k}|=\sum_{jk}|O_{j})\delta _{jk}(\,\rho _{k}|=\sum_{j}|O_{j})(\,\rho
_{j}|=\Pi
\end{equation*}%
}. Therefore, the action of $\Pi $ on $\rho \in \mathcal{O}^{\prime }$
involves a projection leading to a state $\rho _{P}$ such that
\begin{equation}
\rho _{P}\circeq \rho \,\Pi =\sum_{j}(\rho |O_{j})(\,\rho _{j}|  \label{2.11}
\end{equation}%
where in $\rho _{P}$ only contains the information that we can obtain from
the observables $|O_{k})\epsilon \mathcal{O}$

\subsection{ Weyl-Wigner-Moyal mapping.}

Let $\Gamma =\mathcal{M}_{2(N+1)}\equiv \mathbb{R}^{2(N+1)}$ be the phase
space.$_{\text{ }}$The functions over $\Gamma $ will be called $f(\phi )$,
where $\phi $ symbolizes the coordinates of $\Gamma $, $\phi
=(q^{1},...,q^{N+1},p_{q}^{1},...,p_{q}^{N+1})$. If we consider the
operators $\widehat{f,}\widehat{g},...\in \widehat{\mathcal{A}}$ and the
candidates to be the corresponding distribution functions $f(\phi ),g(\phi
),....\in \mathcal{A}$, where $\widehat{\mathcal{A}}$ is the quantum algebra
of operators and $\mathcal{A}$ is the classical algebra \ of distribution
functions, the \textit{Wigner transformation} reads (see \cite{Wigner}, \cite%
{Gadella}, \cite{Symb})

\begin{equation}
symb\widehat{f}\circeq f(\phi )=\int \langle q+\Delta |\,\widehat{f}%
\,|q-\Delta \rangle e^{2i\frac{p\Delta }{\hbar }}d^{N+1}\Delta  \label{2.17}
\end{equation}%
We can also introduce the \textit{star product }(see \cite{Bayern}),$,$%
\begin{equation}
\qquad symb(\widehat{f}\;\widehat{g})=symb\,\widehat{f}\ast symb\,\widehat{g}%
=(f\ast g)(\phi )=f(\phi )\exp \left( -\frac{i\hbar }{2}\overleftarrow{%
\partial }_{a}\omega ^{ab}\overrightarrow{\partial }_{b}\right) g(\phi )
\label{2.18}
\end{equation}%
and the \textit{Moyal bracket, }that is, the symbol corresponding to the
quantum commutator
\begin{equation}
\{f,g\}_{mb}=\frac{1}{i\hbar }(f\ast g-g\ast f)=symb\left( \frac{1}{i\hbar }%
[f,g]\right)  \label{2.19}
\end{equation}%
It can be proved that (see \cite{Wigner})
\begin{equation}
(f\ast g)(\phi )=f(\phi )g(\phi )+0(\hbar ),\text{ }\{f,g\}_{mb}=\{f,g%
\}_{pb}+0(\hbar ^{2})  \label{2.20}
\end{equation}%
To define the inverse $symb^{-1}$, we will use the \textit{symmetrical} or
\textit{Weyl} ordering prescription, namely,
\begin{equation}
symb^{-1}[q^{i}(\phi ),p^{j}(\phi )]\circeq \frac{1}{2}\left( \widehat{q}^{i}%
\widehat{p}^{j}+\widehat{p}^{j}\widehat{q}^{i}\right)  \label{2.21}
\end{equation}%
Therefore, by means of the transformations $symb$ and $symb^{-1}$, we have
defined an isomorphism between the quantum algebra $\widehat{\mathcal{A}}$
and the \textquotedblleft classical-like\textquotedblright\ algebra $%
\mathcal{A}_{q}$,
\begin{equation}
symb^{-1}:\mathcal{A}_{q}\mathcal{\rightarrow }\widehat{\mathcal{A}},\quad
symb:\widehat{\mathcal{A}}\mathcal{\rightarrow A}_{q}  \label{2.22}
\end{equation}%
The mapping so defined is the \textit{Weyl-Wigner-Moyal symbol}.\footnote{%
When $\hbar \rightarrow 0$, then $\mathcal{A}_{q}\rightarrow \mathcal{A}$,
where $\mathcal{A}$ is the classical algebra of observables over phase space.%
}

The Wigner transformation for states is
\begin{equation}
\rho (\phi )=symb\,\widehat{\rho }=(2\pi \hbar )^{-(N+1)\,}symb_{\text{(for
operators)}}\widehat{\,\rho }  \label{2.23}
\end{equation}%
As it is well known, an important property of the Wigner transformation is
that:
\begin{equation}
\langle \widehat{O}\rangle _{\widehat{\rho }}=(\widehat{\rho }|\widehat{O}%
)=(symb\,\widehat{\rho }\,|\,\,symb\,\widehat{O})=\int d\phi ^{2(N+1)}\rho
(\phi )\,O(\phi )  \label{2.24}
\end{equation}%
This means that the definition of $\widehat{\rho }\in \widehat{\text{ }%
\mathcal{A}^{\prime }}$ as a functional on $\widehat{\mathcal{A}}$ is
equivalent to the definition of $symb\,\rho \in $ $\mathcal{A}_{q}^{\prime }$
as a functional on $\mathcal{A}_{q}$.\ \ \

\section{Decoherence in non-integrable systems}

\subsection{Local CSCO}

This subsection is a short version of the corresponding subsection of paper
\cite{CSF}.\bigskip

a.- In \cite{CSF} we have proved that, when the quantum system is endowed
with a CSCO of $N+1$ observables containing $\widehat{H}$, that defines an
eigenbasis in terms of which the state of the system can be expressed, the
corresponding classical system is \textit{integrable}. In fact, if the CSCO
is $\{\widehat{H},\widehat{G}_{1},...,\widehat{G}$ $_{N}\}$, the Moyal
brackets of its elements are
\begin{equation}
\{G_{I}(\phi ),G_{J}(\phi )\}_{mb}=symb\left( \frac{1}{i\hbar }[\widehat{G}%
_{I},\widehat{G}_{J}]\right) =0  \label{3.1}
\end{equation}%
where $I,$ $J=0,1,...,N$, $\widehat{G}_{0}=\widehat{H}$, and $\phi \in
\mathcal{M}\equiv \mathbb{R}^{2(N+1)}$. Then, when $\hbar \rightarrow 0$,
from Eq. (\ref{2.20}) we know that
\begin{equation}
\{G_{I}(\phi ),G_{J}(\phi )\}_{pb}=0  \label{3.2}
\end{equation}%
Thus, since $H(\phi )=G_{0}(\phi )$, the set $\{G_{I}(\phi )\}$ is a
complete set of $N+1$ constants of motion in involution, \textit{globally}
defined all over $\mathcal{M}$; as a consequence, the system is \textit{%
integrable}.\bigskip

b.- We have also proved (see \cite{CSF}) that, when the CSCO has $A+1<N+1$
observables, a local CSCO $\{\widehat{H},\widehat{G}_{1},...,\widehat{G}_{A},%
\widehat{O}_{i(A+1)},...,\widehat{O}_{iN}\}$ can be defined for a maximal
domain \textit{\ }$\mathcal{D}_{\phi _{i}}$\textit{\ }around any point $\phi
_{i}\in \Gamma \equiv \mathbb{R}^{2(N+1)}$, where $\Gamma $ is the phase
space of the system. In this case the system is \textit{non-integrable}.

In order to prove this assertion, we have to recall the \textit{Carath\`{e}%
odory-Jacobi theorem} (see \cite{AM}, theorem 16.29) according to which,
when a system with $N+1$ degrees of freedom has $A+1$ global constants of
motion in involution $\{G_{0}(\phi ),G_{1}(\phi ),...,G_{A}(\phi )\}$, then $%
N-A$ local constants of motion in involution $\{A_{i(A+1)}(\phi
),...,A_{iN}(\phi )\}$ can be defined in a maximal domain $\mathcal{D}_{\phi
_{i}}$\textit{\ }around $\phi _{i}$, for any $\phi _{i}\in \Gamma \equiv
\mathbb{R}^{2(N+1)}$ (see also section 3.2 below).

Let us consider the particular case of a classical system with $N+1$ degrees
of freedom, and whose only global constant of motion (for simplicity) is the
Hamiltonian $H(\phi )$. The Carath\`{e}odory-Jacobi theorem states that, in
this case, the system has $N$ local constants of motion $A_{iI}(\phi )$,
with $I=0,...,N$, in the maximal domain $\mathcal{D}_{\phi _{i}}$\textit{\ }%
around $\phi _{i}$, for any $\phi _{i}\in \Gamma $.

If we want to translate these phase space functions into the quantum
language, we have to apply the transformation $symb^{-1}$; this can be done
in the case of the Hamiltonian, $\widehat{H}=symb^{-1}H(\phi )$, but not in
the case of the $A_{iI}(\phi )$ because they are defined in a maximal domain
$\mathcal{D}_{\phi _{i}}\subset \Gamma $ and the Weyl-Wigner-Moyal mapping
can only be applied on phase space functions defined on the whole phase
space $\Gamma $. To solve this problem, we can introduce a positive
partition of the identity (see \cite{Benatti}),

\begin{equation}
1=I(\phi )=\sum_{i}I_{i}(\phi )  \label{3.3}
\end{equation}
where each $I_{i}(\phi )$ is the \textit{characteristic} or \textit{index}
function

\begin{equation}
I_{i}(\phi )=\left\{
\begin{array}{l}
1\text{ if }\phi \in D_{\phi _{i}} \\
0\text{ if }\phi \notin D_{\phi _{i}}%
\end{array}%
\right.  \label{3.4}
\end{equation}%
and $D_{\phi _{i}}\subset \mathcal{D}_{\phi _{i}}$, $D_{\phi _{i}}$ $\cap $ $%
D_{\phi _{j}}=\emptyset $, $\bigcup_{i}D_{\phi _{i}}=\Gamma $. Then we can
define the functions $O_{iI}(\phi )$ as

\begin{equation}
O_{iI}(\phi )=A_{iI}(\phi )\,I_{i}(\phi )  \label{3.5}
\end{equation}
Now the $O_{iI}(\phi )$ are defined for all $\phi \in \Gamma $; so, we can
obtain the corresponding quantum operators as

\begin{equation}
\widehat{O}_{iI}=symb^{-1}O_{iI}(\phi )  \label{3.6}
\end{equation}%
Since the original functions $A_{iI}(\phi )$ are local constants of motion
in the maximal domain $\mathcal{D}_{\phi _{i}}$, they make zero the
corresponding Poisson brackets, with $H$, in such a domain and, a fortiori,
in the non-maximal domain $D_{\phi _{i}}\subset \mathcal{D}_{\phi _{i}}$.
This means that the $O_{iI}(\phi )$ makes zero the corresponding Poisson
brackets in the whole space space $\Gamma $. In fact, for $\phi \in D_{\phi
_{i}}$, because $O_{iI}(\phi )=A_{iI}(\phi )\,$, and trivially for $\phi
\notin D_{\phi _{i}}$. We also know that, in the macroscopic limit $\hbar
\rightarrow 0$, the Poisson brackets can be identified with the Moyal
brackets, that is, the phase space counterpart of the quantum commutator
(see eq. (\ref{2.20})) \footnote{%
Even if these reasoning is only valid in the limit $\hbar \rightarrow 0$ it
is enough for our purposes since essentially we are trying to find classical
limit.}. Therefore, we can guarantee that all the observables of the set $%
\left\{ \widehat{H},\widehat{O}_{iI}\right\} $ commute with each other:
\begin{equation}
\left[ \widehat{H},\widehat{O}_{iI}\right] =0\qquad \left[ \widehat{O}_{iI},%
\widehat{O}_{iJ}\right] =0  \label{3.6.1}
\end{equation}%
for $I,J=1$ to $N$ and in all the $D_{\phi _{i}}$. As a consequence, we will
say that the set $\left\{ \widehat{H},\widehat{O}_{i1},...,\widehat{O}%
_{iN}\right\} $ is the \textit{local CSCO} of $N+1$ observables
corresponding to the domain $D_{\phi _{i}}\subset \Gamma $. If $\widehat{H}$
has a continuous spectrum $0\leq \omega <\infty $, and the $\widehat{O}_{iI}$
a discrete one (just for simplicity) a generic observable $\widehat{O}$ can
be decomposed as
\begin{equation}
\widehat{O}=\sum_{im_{iI}\,m_{iI}^{\prime }}\int_{0}^{\infty }d\omega
\int_{0}^{\infty }d\omega ^{\prime }\,\,\widetilde{O}_{im_{iI}\,m_{iI}^{%
\prime }}(\omega ,\omega ^{\prime })\,|\omega ,m_{iI}\rangle \langle \omega
^{\prime },m_{iI}^{\prime }|  \label{3.7}
\end{equation}%
where the $|\omega ,m_{iI}\rangle =|\omega ,m_{i1},...,m_{iN}\rangle $ are
the eigenvectors of the local CSCO $\left\{ \widehat{H},\widehat{O}%
_{iI}\right\} $ corresponding to $D_{\phi _{i}}$. Since it can be proved
that (see \cite{CSF}), for $i\neq j$,
\begin{equation}
\langle \omega ,m_{iI}\,|\omega ,m_{jI}\rangle =0  \label{3.8}
\end{equation}%
the decomposition of eq. (\ref{3.7}) is orthonormal, and it generalizes the
usual eigen-decomposition of the integrable case to the non-integrable case.
Therefore, any $\widehat{O}_{iI}$ corresponding to the domain $D_{\phi _{i}}$
commutes with any $\widehat{O}_{jI}$ corresponding to the domain $D_{\phi
_{j}}$ with $i\neq j$,\footnote{%
In this paper we have slightly changed the notation of paper \cite{CSF},
because we consider that the present notation is more explicit than the
one.of that paper.}
\begin{equation}
\left[ \widehat{O}_{iI},\widehat{O}_{jJ}\right] =\delta _{ij}\,\delta _{IJ}
\label{3.8.1}
\end{equation}

\subsection{Continuity and differentiability.}

In paper \cite{CL}, we have used a \textquotedblleft bump\textquotedblright\
smooth function $B_{i}(\phi ),$ in each domain $D_{\phi _{i}}$ surrounded by
a frontier zone $F_{\phi _{i}},$such that $D_{i}(\phi )\cup F_{\phi
_{i}}\subset \mathcal{D}_{i}(\phi ),$ and we have defined a new partition of
the identity (compare with (\ref{3.3})),

\begin{equation}
1=I(\phi )=\sum_{i}B_{i}(\phi )
\end{equation}%
where each $B_{i}(\phi )\geq 0$ satisfies (compare with (\ref{3.4}))

\begin{equation}
B_{i}(\phi )=\left\{
\begin{array}{c}
\begin{array}{l}
1\text{ if }\phi \in D_{\phi _{i}} \\
\epsilon \lbrack 0,1]\text{ if }\phi \notin F_{\phi _{i}}%
\end{array}
\\
0\text{ if }\phi \notin D_{\phi _{i}}\cup F_{\phi _{i}}%
\end{array}%
\right.
\end{equation}%
and $F_{\phi _{i}}\subset \mathcal{F}$ =$\bigcup_{i}F_{\phi _{i}}$ is the
union of all the joining zones (see figure 1.A)\footnote{%
Moreover, as we will discuss in section 5, quantum phase space has a
fundamental graininess. Then the width of $\mathcal{F}$ must be of the order
that we will define in that section, i.e. it must contain a box of the size $%
\Delta x\Delta p=\frac{1}{2}\hbar .$}. Then if we change the definition $%
O_{iI}(\phi )=A_{iI}(\phi )\,I_{i}(\phi $ ) (compare (\ref{3.5})) by%
\begin{equation*}
O_{iI}(\phi )=A_{iI}(\phi )\,B_{i}(\phi )
\end{equation*}%
\ we would have smooth connections between $\mathcal{F}$ \ through the
functions $O_{iI}(\phi )$ \footnote{%
In some cases it can be shown that the discontinuities in the boundary zones
introduces a $0(\hbar ^{2})$, which vanishes when $\hbar \rightarrow 0$ and,
therefore, in this cases, the Moyal brackets can be replaced with Poisson
brackets in such a limit (see \cite{CL})}. Namely to work with continuos and
differential functions force us to introduce continuity zones $\mathcal{F}$ $%
\ $\ and functions $B_{i}(\phi )$ in the frontier of the domains $D_{\phi
_{i}}$ (figure1. A). Then we can use $C^{r}-functions$ (and eventually $%
C^{\infty }-functions)$ in the whole treatment (see$.$\cite{CL}$)$ $.$ For
simplicity, up to now, we have not considered these $\mathcal{F-}$zones,
nevertheless we will be forced to use them in section 6 (figure 6.A). \

Another kind of joining zones are used in the decomposition, in small square
boxes, of a "cell" \cite{Omnes}, i. e. the small boxes distributed in the
"boundary of C" in figure 6.1 of the quoted book (see also between eqs.
(6.6) and (6.79) of this book). This figure corresponds to our figure 1.B.
But, as the $D_{\phi _{i}}$ are neither boxes nor cells (that will be
introduce in section 5), $\mathcal{F}$ \ and the "boundary of C" are
completely different concepts.

\begin{figure}\label{f1}
\centering
\includegraphics{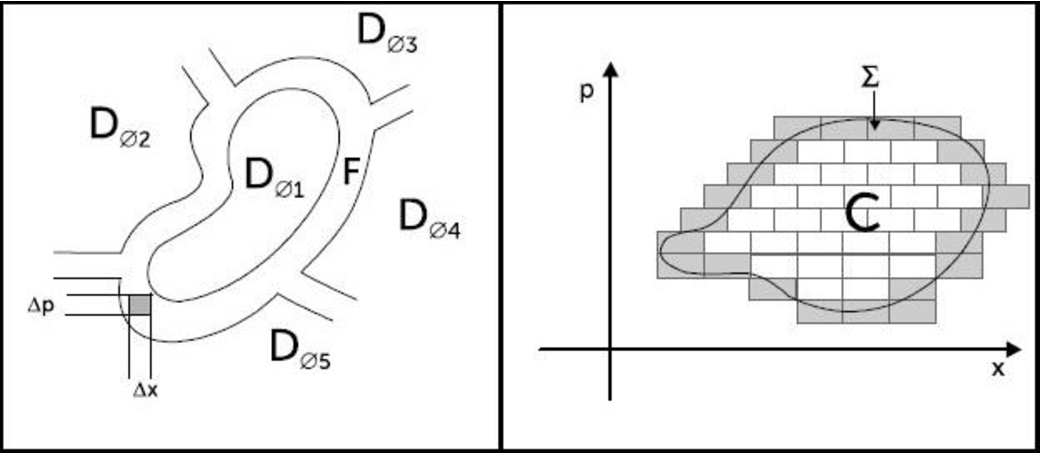}
\caption{\small{Figure 1.A. The domains and the frontier. $\Delta x\Delta p=\frac{1}{2}\hbar$. Figure 1.B
A cell decomposed in small square boxes.}}
\end{figure}

\subsection{Decoherence}

Let us consider a quantum system with a globally defined Hamiltonian $%
\widehat{H}$. In order to complete the CSCO, we can add constants of the
motion locally defined as in the previous subsection. Thus, we have the CSCO
$\left\{ \widehat{H},\widehat{O}_{iI}\right\} $, with $I=1$ to $N$ and $i$
corresponding to all the necessary domains $D_{\phi _{i}}$ obtained from the
partition of the phase space $\Gamma $.\bigskip

a.- In paper \cite{CSF} we have considered the case with continuous and
discrete spectrum for $\widehat{H}$ and for the $\widehat{O}_{iI}$. \ For
the sake of simplicity in this paper we will only consider the continuous
spectrum $0\leq \omega <\infty $ for $\widehat{H}$ and discrete spectra $%
m_{iI}\in \mathbb{N}$ for the $\widehat{O}_{iI}$. Then in the eigenbasis of $%
\widehat{H}$, the elements of any local CSCO can be expressed as (see Eq. (%
\ref{3.7}))
\begin{equation}
\widehat{H}=\sum_{im_{iI}}\,\int_{0}^{\infty }\omega \,|\omega
,m_{iI}\rangle \langle \omega ,m_{iI}|\,d\omega  \label{3.9}
\end{equation}%
\begin{equation}
\widehat{O}_{iJ}=\sum_{im_{iI}}\,\int_{0}^{\infty }m_{iI}\,|\omega
,m_{iI}\rangle \langle \omega ,m_{iI}|\,d\omega  \label{3.10}
\end{equation}%
where $m_{iI}$ is a shorthand for $m_{i1},...,m_{iN}$, and $\sum_{im_{iI}}$
is a shorthand for

$\sum_{i}\sum_{m_{i1}}...\sum_{im_{iN}}$.

With this notation,
\begin{equation}
\widehat{H}\,|\omega ,m_{iI}\rangle =\omega \,|\omega ,m_{iI}\rangle ,\qquad
\widehat{O}_{iI}\,|\omega ,m_{iI}\rangle =m_{iI}\,|\omega ,m_{iI}\rangle
\label{3.13}
\end{equation}%
where the set of vectors $\left\{ |\omega ,m_{iI}\rangle \right\} $, with $%
I=1$ to $N$ and $i$ corresponding to all the domain $D_{\phi _{i}}$, is an
orthonormal basis (see Eq. (\ref{3.8})), i. e.:
\begin{equation}
\langle \omega ,m_{iI}|\,\omega ^{\prime },m_{iI}^{\prime }\rangle =\delta
(\omega -\omega ^{\prime })\,\delta _{m_{iI\,}m_{iI}^{\prime }}\,
\label{3.14}
\end{equation}%
\medskip

b.- Also in the orthonormal basis $\left\{ |\omega ,m_{iI}\rangle \right\} $%
, a generic observable reads (see Eq. (\ref{3.7}))
\begin{equation}
\widehat{O}=\sum_{im_{iI}\,m_{iI}^{\prime }}\int_{0}^{\infty }d\omega
\int_{0}^{\infty }d\omega ^{\prime }\widetilde{O}_{im_{iI}\,m_{iI}^{\prime
}}(\omega ,\omega ^{\prime })\,|\omega ,m_{iI}\rangle \langle \omega
^{\prime },m_{iI}^{\prime }|  \label{3.15}
\end{equation}%
where $\widetilde{O}_{im_{iI}\,m_{iI}^{\prime }}(\omega ,\omega ^{\prime })$
is a generic \textit{kernel} or \textit{distribution} in $\omega ,$ $\omega
^{\prime }$. As in paper \cite{CSF}, we will restrict the set of observables
(i.e. we make a projection like those of section 2.2 namely a generalized
coarse-graining) by only considering the van Hove observables (see \cite{vH}%
) such that \footnote{%
In papers \cite{QCh} we have shown that this choice does not diminish the
physical generality of the model.}
\begin{equation}
\widetilde{O}_{im_{iI}\,m_{iI}^{\prime }}(\omega ,\omega ^{\prime
})=O_{im_{iI}\,m_{iI}^{\prime }}(\omega )\,\delta (\omega -\omega ^{\prime
})+O_{im_{iI}\,m_{iI}^{\prime }}(\omega ,\omega ^{\prime })  \label{3.16}
\end{equation}%
The first term in the r.h.s. of Eq. (\ref{3.16}) is the \textit{singular term%
} and the second one is the \textit{regular term} since the $%
O_{im_{iI}\,m_{iI}^{\prime }}(\omega ,\omega ^{\prime })$ are
\textquotedblleft regular\textquotedblright\ , i. e. $\mathbb{L}_{2}$,
functions of the variable $\omega $ - $\omega ^{\prime }.$ Then we will call
$\widehat{\mathcal{O}}$ the subspace of observable, of our algebra $\widehat{%
\mathcal{A}},$ with these characteristics. Moreover we can define a
projector $\Pi $, as those of section 2.2, that projects on $\widehat{%
\mathcal{O}}\mathcal{.}$ This projection will be our generalized coarse
graining.

Therefore, the observables will read
\begin{eqnarray}
\widehat{O} &=&\sum_{im_{iI}\,m_{iI}^{\prime }}\int_{0}^{\infty }d\omega
O_{im_{iI}\,m_{iI}^{\prime }}(\omega )\,|\omega ,m_{iI}\rangle \langle
\omega ,m_{iI}^{\prime }|+  \notag \\
&&\sum_{im_{iI}\,m_{iI}^{\prime }}\int_{0}^{\infty }d\omega \int_{0}^{\infty
}d\omega ^{\prime }O_{im_{iI}\,m_{iI}^{\prime }}(\omega ,\omega ^{\prime
})\,|\omega ,m_{iI}\rangle \langle \omega ^{\prime },m_{iI}^{\prime }|
\label{3.17}
\end{eqnarray}%
Since the observables are the self-adjoint operators of the algebra, $%
\widehat{O}^{\dagger }=\widehat{O}$, they belong to a space \,\,\,\,\,\,\,\, $\widehat{%
\mathcal{O}}\subset $ $\widehat{\mathcal{A}\text{ }}$ whose basis $\{|\omega
,m_{iI},m_{iI}^{\prime }),|\omega ,\omega ^{\prime },m_{iI},m_{iI}^{\prime
})\}$ is defined as
\begin{equation}
|\omega ,m_{iI},m_{iI}^{\prime })\circeq |\omega ,m_{iI}\rangle \langle
\omega ,m_{iI}^{\prime }|,\qquad |\omega ,\omega ^{\prime
},m_{iI},m_{iI}^{\prime })\circeq |\omega ,m_{iI}\rangle \langle \omega
^{\prime },m_{iI}^{\prime }|  \label{3.18}
\end{equation}%
\medskip

c.- The states belong to a convex set included in the dual of the space $%
\widehat{\mathcal{O}}$, $\widehat{\rho }\in \widehat{\mathcal{S}}\subset
\widehat{\mathcal{O}^{\prime }}$. The basis of $\widehat{\mathcal{O}^{\prime
}}$ is $\{(\omega ,m_{iI},m_{iI}^{\prime }|,(\omega ,\omega ^{\prime
},m_{iI},m_{iI}^{\prime }|\}$, whose elements are defined as functionals by
the equations
\begin{eqnarray}
(\omega ,m_{iI},m_{iI}^{\prime }\,|\,\eta ,n_{iI},n_{iI}^{\prime }) &\circeq
&\delta (\omega -\eta )\,\delta _{m_{iI\,}n_{iI}}\,\delta _{m_{iI}^{\prime
}\,n_{iI}^{^{\prime }}}\,  \notag \\
(\omega ,\omega ^{\prime },m_{iI},m_{iI}^{\prime }\,|\,\eta ,\eta ^{\prime
},n_{iI},n_{iI}^{\prime }) &\circeq &\delta (\omega -\eta )\,\delta (\omega
^{\prime }-\eta ^{\prime })\,\delta _{m_{iI\,}n_{iI}}\,\delta
_{m_{iI}^{\prime }\,n_{iI}^{^{\prime }}}  \notag \\
(\omega ,m_{iI},m_{iI}^{\prime }\,|\,\eta ,\eta ^{\prime
},n_{iI},n_{iI}^{\prime }) &\circeq &0  \label{3.19}
\end{eqnarray}
and the remaining $(\bullet \,\,|\bullet )$ are zero. Then, a generic state
reads

\begin{eqnarray}
\widehat{\rho } &=&\sum_{im_{iI}\,m_{iI}^{\prime }}\int_{0}^{\infty }d\omega
\overline{\rho _{im_{iI}\,m_{iI}^{\prime }}(\omega )}\,(\omega
,m_{iI},m_{iI}^{\prime }|+  \notag \\
&&\sum_{im_{iI}\,m_{iI}^{\prime }}\int_{0}^{\infty }d\omega \int_{0}^{\infty
}d\omega ^{\prime }\overline{\rho _{im_{iI}\,m_{iI}^{\prime }}(\omega
,\omega ^{\prime })}\,(\omega ,\omega ^{\prime },m_{iI},m_{iI}^{\prime }|
\label{3.20}
\end{eqnarray}%
where the functions $\overline{\rho _{im_{iI}\,m_{iI}^{\prime }}(\omega
,\omega ^{\prime })}$ are \textquotedblleft regular\textquotedblright , i.e.
$\mathbb{L}_{2}$\ functions of the variable $\omega $ - $\omega ^{\prime }$.
We also require that $\widehat{\rho }^{\dagger }=\widehat{\rho }$, i.e.,
\begin{equation}
\overline{\rho _{im_{iI}\,m_{iI}^{\prime }}(\omega ,\omega ^{\prime })}=\rho
_{im_{iI}^{\prime }\,m_{iI}}(\omega ^{\prime },\omega )  \label{3.21}
\end{equation}%
and that the $\rho _{im_{iI}\,m_{iI}}(\omega ,\omega )\circeq \rho
_{im_{iI}}(\omega )$ would be real and non-negative, satisfying the total
probability condition,
\begin{equation}
\rho _{im_{iI}}(\omega )\geq 0,\quad \text{ }tr\widehat{\rho }=(\widehat{%
\rho }|\widehat{I})=\sum_{im_{iI}\,}\int_{0}^{\infty }d\omega \rho
_{im_{iI}}(\omega )=1  \label{3.22}
\end{equation}%
where $\widehat{I}=\sum_{im_{iI}\,}\int_{0}^{\infty }d\omega |\omega
,m_{iI}\rangle \langle \omega ,m_{iI}|$ is the identity operator in $%
\widehat{\mathcal{O}}$.\bigskip

d.- On the basis of these characterizations, the expectation value of any
observable $\widehat{O}\in \widehat{\mathcal{O}}$ in the state $\widehat{%
\rho }(t)\in \widehat{\mathcal{S}}$ can be computed as

\begin{equation*}
\langle \widehat{O}\rangle _{\widehat{\rho }(t)}=(\widehat{\rho }(t)|%
\widehat{O})=\sum_{im_{iI}\,m_{iI}^{\prime }}\int_{0}^{\infty }d\omega
\overline{\rho _{im_{iI}\,m_{iI}^{\prime }}(\omega )}\,O_{im_{iI}\,m_{iI}^{%
\prime }}(\omega )+
\end{equation*}
\begin{equation}
\sum_{im_{iI}\,m_{iI}^{\prime }}\int_{0}^{\infty }d\omega \int_{0}^{\infty
}d\omega ^{\prime }\overline{\rho _{im_{iI}\,m_{iI}^{\prime }}(\omega
,\omega ^{\prime })}\,\,e^{i(\omega -\omega ^{\prime })t/\hbar
}\,O_{im_{iI}\,m_{iI}^{\prime }}(\omega ,\omega ^{\prime })  \label{3.23}
\end{equation}
The requirement of \textquotedblleft regularity\textquotedblright,in
variables $\omega -\omega ^{\prime },$ for the involved functions, i. e. $%
O_{im_{iI}\,m_{iI}^{\prime }}(\omega ,\omega ^{\prime })$ $\in $ $%
\mathbb{L}_{2}$ and $\overline{\rho _{im_{iI}\,m_{iI}^{\prime }}(\omega
,\omega ^{\prime })}$ $\in$ $\mathbb{L}_{2},$. As a consequence of Schwartz inequality, it means that
\,\,\,\,\,\,\,\,\,\,\,\,\,\,\,\,\,\,\,\,\,\,\,\,\,\,\,\,\,\,
$\overline{\rho _{im_{iI}\,m_{iI}^{\prime
}}(\omega ,\omega ^{\prime })}\,O_{im_{iI}\,m_{iI}^{\prime }}(\omega ,\omega
^{\prime })\in \mathbb{L}^{1}$ in the variable $\nu =\omega -\omega ^{\prime
}$, a property that we will use below.\medskip

Now, for reasons that will be clear further on, it is convenient to choose a
new basis \{$|\omega ,p_{iI})\}$ that diagonalize the $m$-variables of $\rho
$ (of eq. (\ref{3.21})), for the case $\omega =\omega ^{\prime },$ through a
unitary matrix $U$, which performs the transformation
\begin{equation}
\rho _{im_{iI}\,m_{iI}^{\prime }}(\omega )\rightarrow \rho
_{ip_{iI}\,p_{iI}^{\prime }}(\omega )\,\delta _{p_{iI}\,p_{iI}^{\prime
}}\circeq \rho _{ip_{iI}}(\omega )\,  \label{3.27}
\end{equation}
Such transformation defines the new orthonormal basis $\left\{ |\omega
,p_{iI}\rangle \right\} $, where $p_{iI}$ is a shorthand for $%
p_{i1},...,p_{iN}$, and $p_{iI}\in \mathbb{N}$. This basis corresponds to a
new local CSCO $\left\{ \widehat{H},\widehat{P}_{iI}\right\} $. Therefore,
in each $D_{\phi _{i}}$ we can deduce, from the equations (\ref{3.23}) and (%
\ref{3.27}), that the basis $\left\{ |\omega ,p_{iI}\rangle \right\} $
corresponds to the basis of observables. i. e. $\{|\omega ,p_{iI}),|\omega
,\omega ^{\prime },p_{iI},p_{iI}^{\prime })\}$, defined as in Eq. (\ref{3.18}%
) but with the indices $p$ instead of $m$, and also to the corresponding
basis for the states is $\{(\omega ,p_{iI}|,(\omega ,\omega ^{\prime
},p_{iI},p_{iI}^{\prime }|\}$.\bigskip

Then when the observables $\widehat{P}_{iI}$ have discrete spectra, in the
new basis the van Hove observables of our algebra $\widehat{\mathcal{A}}$
will read
\begin{equation*}
\widehat{O}=\sum_{ip_{iI}\,}\int_{0}^{\infty }d\omega O_{ip_{iI}}(\omega
)\,|\omega ,p_{iI})+
\end{equation*}%
\begin{equation}
\sum_{ip_{iI}\,p_{iI}^{\prime }}\int_{0}^{\infty }d\omega \int_{0}^{\infty
}d\omega ^{\prime }O_{ip_{iI}\,p_{iI}^{\prime }}(\omega ,\omega ^{\prime
})\,|\omega ,\omega ^{\prime },p_{iI},p_{iI}^{\prime })  \label{3.28}
\end{equation}%
where the first term of the r.h.s is the \textit{singular part} and the
second terms the \textit{regular part} of $\widehat{O}$. The states, in
turn, will have the following form
\begin{eqnarray}
\widehat{\rho } &=&\sum_{ip_{iI}\,}\int_{0}^{\infty }d\omega \overline{\rho
_{ip_{iI}}(\omega )}\,(\omega ,p_{iI}|+  \notag \\
&&\sum_{ip_{iI}\,p_{iI}^{\prime }}\int_{0}^{\infty }d\omega \int_{0}^{\infty
}d\omega ^{\prime }\overline{\rho _{ip_{iI}\,p_{iI}^{\prime }}(\omega
,\omega ^{\prime })}\,(\omega ,\omega ^{\prime },p_{iI},p_{iI}^{\prime }|
\label{3.29}
\end{eqnarray}%
where, again, the first term of the r.h.s. is the \textit{singular part} and
the second one is the \textit{regular part} of $\widehat{\rho}$.

\bigskip From the last two equations we have%
\begin{equation*}
(\widehat{\rho (t)}|\widehat{O})=\sum_{ip_{iI}\,}\int_{0}^{\infty }d\omega
\overline{\rho _{ip_{iI}}(\omega )}\,O_{ip_{iI}}(\omega )\,+
\end{equation*}%
\begin{equation*}
\sum_{ip_{iI}\,p_{iI}^{\prime }}\int_{0}^{\infty }d\omega \int_{0}^{\infty
}d\omega ^{\prime }\overline{\rho _{ip_{iI}\,p_{iI}^{\prime }}(\omega
,\omega ^{\prime })}\,\,e^{i(\omega -\omega ^{\prime })t/\hbar
}\,O_{ip_{iI}\,p_{iI}^{\prime }}(\omega ,\omega ^{\prime })\,
\end{equation*}%
Then we can make the Riemann-Lebesgue limit to $(\widehat{\rho }|\widehat{O}%
) $ since from the Schwartz inequality \,\,\,\,\,\,\,\,\,\,\,\,\,\, $O_{ip_{iI}\,p_{iI}^{\prime }}(\omega
,\omega ^{\prime })\overline{\rho _{ip_{iI}\,p_{iI}^{\prime }}(\omega
,\omega ^{\prime })}\in \mathbb{L}_{1}$ in $\nu =$($\omega -\omega
^{\prime })\,$\ the regular part vanishes and only the singular part remains:

\begin{equation}
W-\lim_{t\rightarrow \infty }\widehat{\rho }(t)=\sum_{ip_{iI}\,}\int_{0}^{%
\infty }d\omega \overline{\rho _{ip_{iI}}(\omega )}\,(\omega ,p_{iI}|=%
\widehat{\rho }_{\ast }  \label{3.30}
\end{equation}%
and we have decoherence in all the variables $(\omega ,p_{iI})$\medskip .

Here we have considered the case of observables $\widehat{P}_{iI}$ with
discrete spectra; the case of $\widehat{P}_{iI}$ with continuos spectra is
very similar (see \cite{CSF}).

\subsection{Comment}

A comment is in order: Usually decoherence is studied in the case of open
system surrounded by an environment, up to the point that some people
believe that decoherence takes place in open systems. But also several
authors have introduced, for different reasons, decoherence formalisms for
closed system (\cite{dioisi} - \cite{sicardi}). Related with the method used
in this paper two important examples are given:

1- In paper \cite{connel}, where a system that decoheres at high energy at
the Hamiltonian basis is studied, and

2.-In paper \cite{casati3}, where complexity produces decoherence in a
closed triangular box (in what we could call a Sinai-Young model).

Also we have developed our own theory for decoherence of closed systems, SID
(see \cite{ref19} - \cite{castag4}). In paper \cite{castag5} we show how our
formalism explains the decoherence of the Sinai-Young model above. Recently
it has been shown that also the gravitational field produces decoherence in
the Hamiltonian basis \cite{GP}.

\section{\protect\bigskip The classical statistical limit}

In order to obtain the classical statistical limit, it is necessary to
compute the Wigner transformation of observables and states. For
simplicity and symmetry we will consider all the variables $(\omega ,p_{iI})$ continuous in this section. If we do this substitution, Eq. (\ref{3.29}), reads%
\begin{equation}\label{eq4.1}
\widehat{\rho }(t)=\sum_{i}\int_{p_{iI}}dp_{iI}^{N}\,\int_{0}^{\infty
}d\omega \,\overline{\rho _{i}(\omega ,p_{iI})}\,(\omega ,p_{iI}|+
\end{equation}
\begin{equation}
\sum_{i}\int_{p_{iI}}dp_{iI}^{N}\int_{p_{iI}^{\prime }}dp_{iI}^{\prime
N}\,\int_{0}^{\infty }d\omega \int_{0}^{\infty }d\omega ^{\prime }\,%
\overline{\rho _{i}(\omega ,\omega ^{\prime },p_{iI},p_{iI}^{\prime })}%
\,\,e^{i(\omega -\omega ^{\prime })t/\hbar }\,(\omega ,\omega ^{\prime
},p_{iI},p_{iI}^{\prime }|  \label{4.3}
\end{equation}%
Therefore, Eq. (\ref{3.30}) can be written as
\begin{equation}
W-\lim_{t\rightarrow \infty }\widehat{\rho }(t)=\widehat{\rho }_{\ast
}=\sum_{i}\int_{p_{iI}}dp_{iI}^{N}\,\int_{0}^{\infty }d\omega \overline{\rho
_{i}(\omega ,p_{iI})}\,(\omega ,p_{iI}|  \label{4.4}
\end{equation}%
where $\widehat{\rho }_{\ast }$ is simply the singular component of $%
\widehat{\rho }(t)$, where the regular part has vanished as a consequence of
the Riemann-Lebesgue theorem.

Now, the task is to find the classical distribution $\rho _{\ast }(\phi )$
resulting from the Wigner transformation of $\widehat{\rho }_{\ast }$ in the
limit $\hbar \rightarrow 0$,
\begin{equation}
\rho _{\ast }(\phi )=symb\,\widehat{\rho }_{\ast }  \label{4.10}
\end{equation}%
where
\begin{equation}
\rho _{\ast }(\phi )=symb\,\widehat{\rho }_{\ast
}=\sum_{i}\int_{p_{iI}}dp_{iI}^{N}\int_{0}^{\infty }d\omega \,\overline{\rho
_{i}(\omega ,p_{iI})}\,\,\,symb\,(\omega ,p_{iI}\,|  \label{4.12}
\end{equation}%
So, the problem is reduced to compute $symb\,(\omega ,p_{iI}\,|$.\bigskip

As it is well known, in its traditional form the Wigner transformation
yields the correct expectation value of any observable in a given state when
we are dealing with regular functions (see Eq. (\ref{2.24})). In previous
papers (\cite{CSF}, \cite{Cast-Gadella 2006}) we have extended the Wigner
transformation to singular functions in order to use it in functions like $%
(\omega ,p_{iI}\,|$. Here we will briefly resume the results of these papers
in two steps: first, we will consider the transformation of observables and,
second, we will study the transformation of states.

\subsection{Transformation of observables}

As we have seen (see Eq.(\ref{3.28})), our van Hove observables $\widehat{O}%
\in \widehat{\mathcal{O}}$ have a singular part, i. e. $\widehat{O}_{S},$
and a regular part, i.e. $\widehat{O.}_{R}$. We will direct our attention to
the singular operators $\widehat{O}_{S}$, since the regular operators $%
\widehat{O}_{R}$ \textquotedblleft disappear\textquotedblright\ from the
expectation values after decoherence, as explained in Section 2.3. $\widehat{%
O}_{S}$ reads:
\begin{equation}
\widehat{O}_{S}=\sum_{i}\int_{p_{iI}}dp_{iI}^{N}\,\int_{0}^{\infty }d\omega
O_{i}(\omega ,p_{iI})\,|\omega ,p_{iI})  \label{4.13}
\end{equation}%
Then, the Wigner transformation of $\widehat{O}_{S}$ can be computed as
\begin{equation}
O_{S}(\phi )=symb\,\widehat{O}_{S}  \label{4.16}
\end{equation}%
where
\begin{equation}
O_{S}(\phi )=symb\,\widehat{O}_{S}=\sum_{i}\int_{p_{iI}}dp_{iI}^{N}\,%
\int_{0}^{\infty }d\omega \,O_{i}(\omega ,p_{iI})\,\,\,symb\,|\omega ,p_{iI})
\label{4.18'}
\end{equation}%
Now if we consider that the functions $O_{i}(\omega ,p_{iI})$ are
polynomials of functions of a certain space where the polynomials are dense
it can be probed that%
\begin{equation*}
\widehat{O}_{S}=\sum_{i}O_{\phi _{i}}(\widehat{H},\widehat{P}%
_{iI})\,=\sum_{i}\widehat{O_{S\phi _{i}}}
\end{equation*}%
where $\widehat{O_{S\phi _{i}}}=O_{S\phi _{i}}(\widehat{H},\widehat{P}%
_{iI}), $ and where $symb\widehat{O_{S\phi _{i}}}=symbO_{S\phi _{i}}(%
\widehat{H},\widehat{P}_{iI})=O_{S\phi _{i}}(H(\phi ),P_{iI}(\phi ))+O(\hbar
^{2})$. Then if $O_{S\phi _{i}}(H(\phi ),P_{iI}(\phi ))=\delta (\omega
-\omega ^{\prime })\,\delta ^{N}(p_{iI}-p_{iI}^{\prime })$ we have (see
paper \cite{CL} for details) that the function $symb\,|\omega ,p_{iI})$ in
the limit $\frac{\hbar }{S}\rightarrow 0,$ is
\begin{equation}
symb\,|\omega ,p_{iI})=\delta (H(\phi )-\omega )\,\delta ^{N}(P_{iI}(\phi
)-p_{iI})  \label{4.19}
\end{equation}%
where $H(\phi )=symb\widehat{H\text{ }}$ and $P_{iI}(\phi )=symb\widehat{%
P_{iI}\text{ }}$

\subsection{Transformation of states}

As in papers \cite{CSF} and \cite{Cast-Gadella 2006}, in order to compute
the $symb$ $(\omega ,p_{iI}|,$ we will define the Wigner transformation of
the singular operator $\widehat{\rho }_{S}=\widehat{\rho }_{\ast }$ on the
base of the only reasonable requirement that such a transformation would
lead to the correct expectation value of any observable. Then we must
postulate that it is (see Eq. (\ref{2.24})),
\begin{equation}
(symb\,\widehat{\rho }_{S}\,|\,\,symb\,\widehat{O}_{S})\circeq (\widehat{%
\rho }_{S}\,|\,\widehat{O}_{S})  \label{4.21}
\end{equation}%
These equations must also hold in the particular case in which $\widehat{O}%
_{S}=|\omega ^{\prime },p_{iI}^{\prime })$, $\widehat{\rho }_{S}=(\omega
,p_{iI}|$, for some $D_{\phi _{i}}$ (see Eq.(\ref{3.8})) i. e.:

\begin{equation}
(symb(\omega ,p_{iI}|\,\,|\,\,symb\,|\omega ^{\prime },p_{iI}^{\prime
}))=(\omega ,p_{iI}\,|\,\omega ^{\prime },p_{iI}^{\prime })  \label{4.25'}
\end{equation}%
and all the remaining cross terms are zero for any domain $D_{\phi _{j}}$,
with $j\neq i$. But from Eq. (\ref{4.19}) we know how to compute $%
symb\,|\omega ^{\prime },p_{iI}^{\prime })$. Moreover, from the definition
of the cobasis (see Eq. (\ref{3.19})) we know that
\begin{equation}
(\omega ,p_{iI}\,|\,\omega ^{\prime },p_{iI}^{\prime })=\delta (\omega
-\omega ^{\prime })\,\delta ^{N}(p_{iI}-p_{iI}^{\prime })  \label{4.27'}
\end{equation}%
Therefore in the limit $\frac{\hbar }{S}\rightarrow 0$ we have,

\begin{equation}
(symb(\omega ,p_{iI}|\,\,|\,\,\delta (H(\phi )-\omega ^{\prime })\,\delta
^{N}(P_{iI}(\phi )-p_{iI}^{\prime }))=\delta (\omega -\omega ^{\prime
})\,\delta ^{N}(p_{iI}-p_{iI}^{\prime })  \label{4.29'}
\end{equation}%
Then in paper \cite{CSF} we have proved that (always in the $\frac{\hbar }{S}%
\rightarrow 0$ limit)
\begin{equation}
symb(\omega ,p_{iI}|=\frac{\delta (H(\phi )-\omega )\,\delta
^{N}(P_{iI}(\phi )-p_{iI})}{C_{i}(H,P_{iI})}  \label{4.30}
\end{equation}%
where $C_{i}(H,P_{iI})$ is the configuration volume of the region $\Gamma
_{H,P_{iI}}\cap D_{\phi _{i}}$, being $\Gamma _{H,P_{iI}}\subset \Gamma $
the hypersurface defined by $H=const.$ and $P_{iI}=const.$ In this way we
have obtained the $symb$ of $\,|\omega ,p_{iI})$ and $(\omega ,p_{iI}|$ so
the classical statistical limit is completed.

\subsection{Convergence in phase space}

Finally, we can introduce the results of Eq. (\ref{4.30}) into Eq. (\ref%
{4.12}), in order to obtain the classical distribution $\rho (\phi )$:%
\begin{equation}
\rho _{\ast }(\phi )=\rho _{S}(\phi
)=\sum_{i}\int_{p_{iI}}dp_{iI}^{N}\,\int_{0}^{\infty }d\omega \,\frac{%
\overline{\rho _{i}(\omega ,p_{iI})}}{C_{i}(H,P_{iI})}\,\,\,\delta (H(\phi
)-\omega )\,\delta ^{N}(P_{iI}(\phi )-p_{iI})  \label{4.33'}
\end{equation}%
As a consequence, the Wigner transformation of the limits of Eq. (\ref{3.30}%
) can be written as%
\begin{equation*}
W-\lim_{t\rightarrow \infty }\rho (\phi ,t)=\rho _{S}(\phi )=\rho _{\ast
}(\phi )=
\end{equation*}%
\begin{equation}
\sum_{i}\int_{p_{iI}}dp_{iI}^{N}\,\int_{0}^{\infty }d\omega \,\frac{%
\overline{\rho _{i}(\omega ,p_{iI})}}{C_{i}(H,P_{iI})}\,\,\,\delta (H(\phi
)-\omega )\,\delta ^{N}(P_{iI}(\phi )-p_{iI})  \label{4.35'}
\end{equation}

\noindent Remember that all this is only valid in a domain $D_{\phi }$ defined in eq.
(19) and that it would completely change if we change to another domain
through a continuity zone $\mathcal{F}$ of section 3.2.

Then we have obtained a convincing classical limit of the states, that
decomposed as in eq. \ref{4.35'}, it turns out to be sums of states peaked
in the classical hypersurfaces of constant energy, $H(\phi )=\omega ,$ and
where also the other constants of motions are constant, $P_{iI}(\phi
)=p_{iI} $. This is an important step forward, to have obtained these
classical surfaces as a limit of the quantum mechanics formalism. Up to here
chaos has not produced any problem to the CP, even if the system is not
integrable. The real problems will begin in the next section.

\section{Graininess}
This section is devoted to a brief review of the graininess in quantum mechanics and some of its approaches. First we start with the standard approach from the viewpoint of the Kolmogorov-Sinai entropy and its quantum variants. Then we introduce our alternative approach using cells of the phase space that are deformed as the system evolves and where are all physical magnitudes are coarse-grained on domains of minimal size in the sense of the Indetermination Principle.
\subsection{The standard approach}
The two properties of classical mechanics necessary for chaos to occur are a continuous spectrum and a continuous phase space \cite{casati chirikov}. On the other hand, the most quantum systems which present chaotic features in its classical limit have discrete spectrum. In addition, the Correspondence Principle CP implies the transition from quantum to classical mechanics for all phenomena including the chaos. However, by the Indetermination Principle in quantum mechanics we have a discretized, and therefore non-continuous, phase space divided in elementary cells of finite size
$\Delta x\Delta p\geq \hbar$ (per freedom). Then the natural question that arises is: How can we provide a quantum formalism consistent with the Indetermination principle and the CP to explain the emergence of chaos in the classical limit?

This is where the treatment of the graininess arises as a possible answer to the problem. The graininess has several approaches that try to solve the problem without being a threat to the CP. The ``natural" possibility of accomplishing this could be the quantization of chaotic systems, but due the compactness of its phase space the quantization yields discrete energy spectrum. Then the situation does not seem so simple at first glance and one must look for other indicators that somehow capture the main properties related to the continuous spectrum of chaotic systems. The Kolmogorov-Sinai entropy \cite{katok} (KS-entropy) is perhaps the most significant and robust indicator, both in theory and applications. Roughly speaking, one reason why this is so is due one can model the behavior of classical chaotic systems of continuous spectrum from classical discretized models such that the KS-entropies of the continuous system and of the discrete ones tend to coincide for a certain appropriate range. We recall that the KS-entropy assigns measures to bunches of trajectories and computes the Shannon-entropy
per time-step of the ensemble of bunches in the limit of infinitely many
time-steps and the Pesin theorem \cite{mane} links the KS-entropy with the Lyapunov coefficients. For a quantum description of the chaotic systems, we would need a quantum extension of the KS-entropy. There are several non-conmutative candidates \cite{connes, alicki, voiculescu, accardi, zyczkowski} and the presence of a finite time interval where these KS-extensions yield the KS-entropy is considered as the main peculiarity of quantum chaos \cite{casati chirikov}.
Therefore the issue of graininess is intimately related to quantum chaos timescales and must necessarily be compatible with the restriction to these. Three time scales characterizing the classically chaotic quantum motion are distinguished: The relaxation time scale, the random time scale and the logarithmic breaking time. Only for regular classical limits classical and quantum mechanics are expected to overlap over times $t$ such that
\begin{equation}\label{heisenberg time}
t\lesssim t_R \propto \hbar^{-\alpha} \,\,\,\,\,\,\,\, for \,\,some \,\,\,\,\,\,\,\, \alpha>0
\end{equation}
\noindent where $t_R$ is the relaxation time scale which determines the so-called semi-classical regime, i.e. the time scale where the phenomena like the exponential localization and the relaxation can occur. Moreover the discrete spectrum cannot be solved if $t\lesssim t_R$ (see pag. 12 of \cite{casati chirikov}). The breaking time scale (random time scale) $\tau$ is much shorter than $t_R$ and is related to a stronger chaotic property, the exponential instability. Basically, $\tau$ determines the time interval where the wave-packet motion is as random as the classical trajectory and the time for the spreading of the packet is given by
\begin{equation}\label{ehrenfest time}
\tau\sim \ln \frac{q}{h}\propto -\log \hbar
\end{equation}
\noindent where $q$ is the quasiclassical parameter which is of the order of the characteristic value of the action value (see pag. 14 of \cite{casati chirikov}). %Finally, the logarithmic breaking time $\tau$ says that for chaotic classical limits classical and quantum mechanics agree over times $t$ such that
%\begin{equation}\label{logarithmic time}
%t\simeq \tau \propto -\log \hbar
%\end{equation}
The importance of the logarithmic breaking time $\tau$ is that this indicates the typical scaling
for a joint time-classical limit suited to classically chaotic quantum systems. We should mention that some authors, see \cite{casati chirikov}, consider that $\tau$ is a satisfactory resolution between of the apparent contradiction between the CP and the quantum transient (finite-time) given by $t_r$ and the evidence that time and classical limits do not commute. That is,
\begin{equation}\label{double limit}
lim_{|t|\rightarrow \infty} lim_{q\rightarrow \infty}\neq lim_{q\rightarrow \infty} lim_{|t|\rightarrow \infty}
\end{equation}
\noindent where the first order leads to classical chaos and the second one represents a quantum behavior with no chaos at all (see pag. 17 of \cite{casati chirikov}).

Then if we define\footnote{Where $S$ denotes the action.} $q=\frac{S}{\hbar}$ we could claim that the classical statistical limit of the section 4 (i.e. $t\rightarrow\infty,\frac{\hbar}{S}\rightarrow0$) is quite similar to the double limit of the right hand of the eq. \eqref{double limit}. In section 5.4 we will discuss this situation taking into account the graininess and the quantum chaos timescales. In the next two sections we introduce our graininess approach considering phase space cells as the starting point.

\subsection{Our approach: fundamental graininess with cells}

In this section we describe our approach of the graininess. As we mentioned in the introduction the key is to average a point-test-distribution function on minimal rectangular boxes of the phase space. The motivation of this approach lies in the fact that we can obtain a classical limit (and its limitations) searching the trajectories of the rectangular boxes (and later of the cells) we will consider as ``points", integrating the Heisenberg equation, and then studying the deformations of the cells under the motion (as in \cite{Omnes}).

In section 4.3 we have found the hypersurfaces where the classical trajectories
lay. Now we want to find the classical motions in these trajectories. Thus
we need to define the notion of ``\emph{a point that moves}". But in quantum
mechanics there is not such a thing. In fact it is well known that the
commutation relations and its consequence, the indetermination principle,
establishes a fundamental graininess in the ``quantum phase space". Precisely if
we call $\widehat{J}$ and $\widehat{\Theta }$ two generic conjugated
operators (e. g. in our case $\widehat{J}$ will be the constants of the
motion $\widehat{H\text{ }},$ $\widehat{P}_{iI}$ and $\widehat{\Theta }$ the
corresponding configuration operators) we have
\begin{equation}
\lbrack \widehat{\Theta },\widehat{J}]=i\hbar \widehat{I}  \label{C}
\end{equation}%
and therefore
\begin{equation}
\Delta _{\Theta }\Delta _{J}\geq \frac{\hbar }{2}  \label{U}
\end{equation}%
where, from now on, $\Delta _{\Theta }$ and $\Delta _{J}$ are defined as the
variances of some \textit{typical} state $\widehat{\rho}$ the one with the
smallest dimensions we can ``determinate" (in the sense of Ballentine chapter 8
\cite{Ballentine}) in our experiment. With different choices for this $\widehat{\rho}$
we will obtain different ratios $\Delta _{\Theta }/\Delta _{J}$ but the
qualitative results will be the same. Then we will consider that the
rectangular box $\Delta _{\Theta }\Delta _{J}$ of volume $\hbar $ (or the
polyhedral box of volume $\hbar ^{(n+1)}$ in the many dimensions case)
will be the smallest volume that we can determinate with our measurement
apparatus, precisely:
\begin{equation*}
vol\Delta _{\Theta }\Delta _{J}=\hbar \text{ (or eventually }N_{0}\hbar )%
\text{ }
\end{equation*}%
for a phase space of two dimensions or%
\begin{equation*}
vol\prod \Delta _{\Theta }\prod \Delta _{J}=\hbar ^{(n+1)}\text{ (or
eventually }N_{0}\hbar ^{(n+1)})\text{ }
\end{equation*}%
for a phase space of $2(n+1)$ dimensions, where $N_{0}$ is not a very large
natural number (cf. \cite{Omnes}). This is the new feature of the ``quantum
phase space": its graininess and this fact will be the origin of the threat
to the CP \footnote{%
Fundamental graininess appears in many other disguises (see \cite{Peskin},
\cite{Sorkin}, etc.)}.

In Omn\`{e}s book \cite{Omnes} the cells produced by the fundamental
graininess are described in the ($x,p)$ coordinates, using a mathematical
theory, the microlocal analysis, based in the work \cite{SCF}. In our
formalism we will change these ($x,p)$ for the ($J,\Theta )$ coordinates
where $J$ are the constants of the motion and $\Theta $ the corresponding
configuration variables and where the commutation relations (\ref{C}) and
their consequence the indetermination principle (\ref{U}) will play the main
role.

To see how the fundamental graininess works let us consider a closed simply
connected set of a two dimensional phase space that we will call a cell \ $%
C^{T}$, with its continuous boundary $B,$ (figure 1.B, or fig 6.1 of \cite%
{Omnes}). The coordinates ($J,\Theta )$ and a lattice of rectangular boxes $%
\Delta _{\Theta }\Delta _{J}$ (eventually $2(n+1)$ polyhedral boxes) define
the two domains related with $C^{T}$: $\Sigma $, set of boxes that intersect
$B,$ and $C$, the set of the interior rectangular boxes of the cell $C^{T}.$
Volume is well defined in phase space of any dimension while (hyper)
surfaces are not defined, so in order to compare the the size of the
frontier with the size of the interior we can define the adimensional parameter
\begin{equation*}
\Omega =\frac{vol\Sigma }{volC}
\end{equation*}%
It is quite clear that $\Omega \ll 1$ corresponds to a bulky cell while $%
\Omega \gg 1$ corresponds to an elongated and maybe deformed cell. It is
also almost evident that if we want that a cell would somehow represent a
real point it is necessary that $\Omega <1,$ because if $\Omega >1$ the
volume of the interior $C$ is smaller than the volume of the ``frontier" $%
\Sigma ,$ where we do not know for certain if its points belong or not to $%
C^{T}$ since $B\subset \Sigma .$ Thus in the case $\Omega \gg 1$ we
completely lose the notion of real point and the description of the
classical trajectories, as the motion of $C^{T},$ becomes impossible.

Analogously Omn\`{e}s defines semiclassical projectors for each cell and
shows that if $\Omega $ is very large the definition of these projectors
lose all its meaning and the classicality is lost, namely he obtains a
similar conclusion.

In the next section we will consider the cells and their evolution in
several cases and we will estimate their corresponding $\Omega$. From now, in all cases where the quasiclassical parameter $q=\frac{S}{\hbar}$ is finite it should be noted that we mean ``a threat to the CP" to the outside time range of validity of our graininess approach according to the CP and should not be necessarily associated with the emergence of the non-commutative two limits given by the eq. \eqref{double limit}. Furthermore, since the timescales considered in our graininess approach will be finite then there is no way that any of the two limits of eq. \eqref{double limit} appear. All this will be discussed in the next section.

\subsection{The classical trajectories}

Up to this point we have obtained the classical distribution $\rho _{\ast
}(\phi )=\rho _{S}(\phi )$ to which the system converges in phase space.
This distribution defines hypersurfaces $H(\phi )=\omega $, $P_{iI}(\phi
)=p_{iI}$ corresponding to the constant of the motion i.e. our the
\textquotedblleft momentum\textquotedblright\ variables. But such a
distribution does not define the trajectories of ``points" on those
hypersurfaces, i. e., it does not fix definite values for the
\textquotedblleft configuration\textquotedblright\ variables (the variables
canonically conjugated to $H(\phi )$ and $P_{iI}(\phi )$). This is
reasonable to the extent that definite trajectories would violate the
uncertainty principle. In fact we know that, if $\widehat{H}$ and $\widehat{P%
}_{iI}$ have definite values, then the values of the observables that do
non-commute with them will be completely undefined.

As in section 5.2, let
us call, $\widehat{J}$ the \textquotedblleft momentum\textquotedblright\
variables $\widehat{H}$ and $\widehat{P}_{iI}$ (constants of the motion),
and $\widehat{\Theta }$ the corresponding conjugated \textquotedblleft
configuration\textquotedblright\ variables, all of them defined in the
domain $D_{\phi _{i}}$. The equations of motion, in the Heisenberg picture,
read
\begin{equation}
\frac{d\widehat{J}}{dt}=\frac{i}{\hbar }[\widehat{H},\widehat{J}]\qquad
\quad \frac{d\widehat{\Theta }}{dt}=\frac{i}{\hbar }[\widehat{H},\widehat{%
\Theta }]  \label{5.1}
\end{equation}%
where as $[\widehat{H},\widehat{J}]=0$
\begin{equation}
\frac{d\widehat{J}}{dt}=0\qquad \quad \frac{d\widehat{\Theta }}{dt}=\frac{i}{%
\hbar }[\widehat{H},\widehat{\Theta }]  \label{5.2}
\end{equation}
Within the domain $D_{\phi }$ we know that if we can consider the $\widehat{H}$
as a function (or a convergent sum)\ of the $\widehat{J},$ i. e.:
\begin{equation}
\widehat{H}=F(\widehat{J})=\sum_{n}a_{n}\widehat{J}^{n}  \label{5.2"}
\end{equation}
and since $[\widehat{\Theta },\widehat{J}]=i\hbar \widehat{I} $ we have $[\widehat{
\Theta},\widehat{J}^{n}]=in\hbar \widehat{J}^{(n-1)}$ so
\begin{equation*}
\lbrack \widehat{H},\widehat{\Theta }]=\frac{d\widehat{H}}{d\widehat{J}}
\end{equation*}%
where $\widehat{H}$ and $\widehat{J}$ are constant in time, so calling $%
\widehat{V}(0)=\frac{d\widehat{H}}{d\widehat{J}},$ which is another constant
in time, we have%
\begin{equation*}
\widehat{J}(t)=\widehat{J}(0),\text{ \ \ }\widehat{\Theta }(t)=\widehat{%
\Theta }(0)+\widehat{V}(0)t
\end{equation*}%
Then we can make the Wigner transformation from these equations and, since
this transformation is linear, we have%
\begin{equation}
J(\phi ,t)=J(\phi ,0),\text{ \ \ }\Theta (\phi ,t)=\Theta (\phi ,0)+V(\phi
,0)t  \label{WT1}
\end{equation}%
We will use this equation to follow the motion of the boxes and the cells in
the phase space:

Let us first consider a \textit{rectangular} (eventually $2(n+1)$ polyhedral)
moving box of size $\Delta _{\Theta }\Delta _{J}$ with $\Delta _{\Theta
}\Delta _{J}\sim \hbar $ (eventually $\hbar ^{(n+1)}$), that we will
symbolize by a small square in figures 2, 3, 4, and 5 (and just by a point
in the figures 6.A and 6.B) and let us also consider \ the \textit{typical}
point-test-distribution function $symb \ \widehat{\rho}=\rho (\phi)=\rho
(j,\theta ),$ (see under eq. (\ref{U}), also from now on $\phi =(j,\theta )$%
) with support contained in $\Delta _{\Theta }\Delta _{J},$ then \ let us
define the mean values%
\begin{equation*}
\overline{j(t)}=\int_{\Delta _{\Theta }\Delta _{J}}J(j,\theta ,t)\rho
(j,\theta )djd\theta ;\text{ \ }\overline{\theta (t)}=\int_{\Delta _{\Theta
}\Delta _{J}}\Theta (j,\theta ,t)\rho (j,\theta )djd\theta ,
\end{equation*}%
\begin{equation}
\text{ \ }\overline{v(t)}=\int_{\Delta _{\Theta }\Delta _{J}}V(j,\theta
,t)\rho (j,\theta )djd\theta ;  \label{S}
\end{equation}%
where the $\rho (j,\theta )$ is not a function of the time since we are in
the Heisenberg picture and
\begin{equation}
\begin{split}
& J(j,\theta ,t)=symb \ \widehat{J}(t) \\
& \Theta(j,\theta ,t)=symb \ \widehat{\Theta}(t) \\
& V(j,\theta,t)=symb \ \widehat{V}(0)
\end{split}
\end{equation}
Now using eq. (\ref{WT1}) we have%
\begin{equation}
\overline{j}(\phi ,t)=\overline{j}(\phi ,0),\text{ \ \ }\overline{\theta }%
(\phi ,t)=\overline{\theta }(\phi ,0)+\overline{v}(\phi ,0)t  \label{S'}
\end{equation}%
so our minimal rectangular box moves along a classical trajectory of our
system.

Now our rectangular boxes are so small that we can not even consider their
possible deformation. Precisely the Indetermination Principle makes this
deformation merely hypothetical. Thus, from now on, we will consider that
the rectangular boxes \textit{\ are not in motion (and therefore they can
not be deformed} \textit{by motion)} and that they are the most elementary
theoretical fixed notion of a point at $(\overline{j},\overline{\theta })$.
\footnote{The rectangular moving cell defined after the eq. (\ref{WT1}) will be the
only rectangular objects that moves in this paper.} In this way we have
obtained the classical trajectories of theoretical points (i.e. eq.(\ref{S'}))
and we would have completed our quantum to classical limit (apparently CP
is safe up to now).

But remember that the real physical points are not these rectangular boxes
but the cells with $\Omega <1$ that we must also consider, because real
measurement devices cannot see the elementary rectangular boxes but bigger
cells of dimensions far bigger than the Planck ones. In the next examples we
will see what happens with these cells that we will consider as real points:
the cells can be deformed by the motion (while the rectangular boxes always
remain rigid). We will show the interplay of these theoretical points
(boxes) and physical real points (cells) in some examples bellow:

1.- Then, as a first example, let us consider a two dimensional space within
a domain $D_{\phi }$ (much larger than the cell that we will define below)
and let as also consider the system of coordinates $(J,\Theta )$ and the
corresponding trajectories when the Hamiltonian is a linear function, $%
\widehat{H}=a_{0}\widehat{I}+a_{1}\widehat{J},$ Then $\widehat{V}=a_{1}%
\widehat{I}$ so%
\begin{equation*}
J(\phi ,t)=J(\phi ,0),\text{ \ \ }\Theta (\phi ,t)=\Theta (\phi ,0)+a_{1}It
\end{equation*}%
and, with the same reasoning as above the trajectories of the boxes
(theoretical points) are

\begin{equation}
\overline{j}(t)=\overline{j}(0),\text{ \ \ }\overline{\theta }(t)=\overline{%
\theta }(0)+a_{1}t  \label{K}
\end{equation}

\noindent Namely we obtain the figure 2 and we have a uniform translation motion with
constant velocity $\overline{v}[\overline{j}(0)]$ along all the trajectories. Let us then consider two parallel lines with constant velocities $%
\overline{v}[\overline{j}_{1}]=\overline{v}[\overline{j_{2}}],$ thus the
difference of velocities is%
\begin{equation}\label{A}
\overline{v}(\overline{j}_{1})-\overline{v}(\overline{j_{2}})=0
\end{equation}%
\noindent Then if we consider an initial rectangular cell the motion will not
deform the cell. Since there is no deformation of the cell $\Omega $ is
rigid, thus if $\Omega <1$ in the initial cell $\Omega $ will be $<1$ in any
transferred cell. Therefore, in this trivial case the cell will represent a
physical real point moving according to eq. (\ref{K}). Thus in this case we
have completed our classical limit and the CP is safe.

\begin{figure}\label{f2}
\centering
\includegraphics[width=0.7\textwidth]{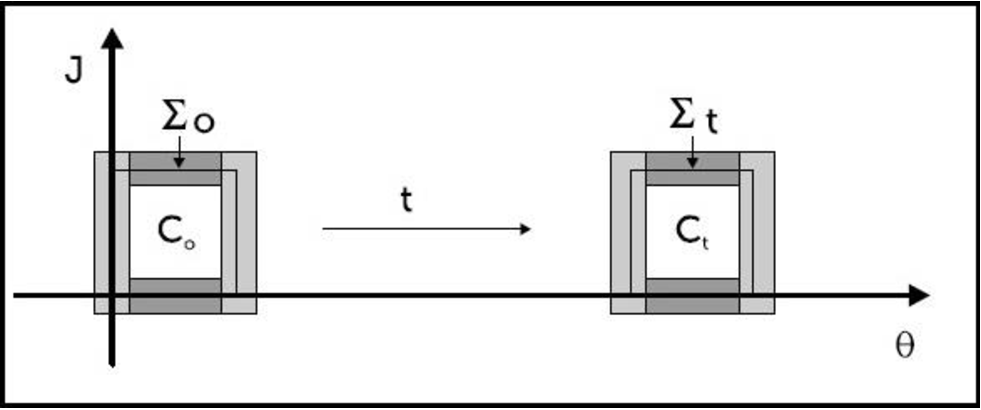}
\caption{\small{Evolution of a cell with constant velocity.}}
\end{figure}

2.- As a further example let us consider the same two dimensional space
within a $D_{\phi }$ and let as consider the system of coordinates $%
(J,\Theta )$ and the corresponding trajectories when $\widehat{H}=a_{0}%
\widehat{I}+a_{1}\widehat{J}+a_{2}\widehat{J}^{2}.$ Then $\widehat{V}=a_{1}%
\widehat{I}+a_{2}\widehat{J}$ so%
\begin{equation*}
J(\phi ,t)=J(\phi ,0),\text{ \ \ }\Theta (\phi,t)=\Theta (\phi
,0)+[a_{1}I+2a_{2}J(\phi ,0)]t
\end{equation*}
and, with the same reasoning as above
\begin{equation*}
\overline{j}(t)=\overline{j}(0),\text{ \ \ }\overline{\theta }(t)=\overline{%
\theta }(0)+[a_{1}+2a_{2}\overline{j}(0)]t
\end{equation*}
Namely we obtain the figure 3 and we have a uniform motion with constant
velocity $\overline{v}[\overline{j}(0)]=a_{1}+2a_{2}\overline{j}(0)$ along
straight lines parallel to the axis $\theta$. That is,
\begin{equation*}
\text{\ \ }\overline{\theta }(t)=\overline{\theta }(0)+\overline{v}[%
\overline{j}(0)]t
\end{equation*}%
Let us then consider two parallel lines with constant velocities $\overline{v%
}(\overline{j}_{1})\neq \overline{v}(\overline{j_{2}}),$ thus the
difference of velocities is
\begin{equation}
\overline{v}(\overline{j}_{1})-\overline{v}(\overline{j_{2}})=2a_{2}(%
\overline{j}_{1}-\overline{j_{2}})=v  \label{B}
\end{equation}
Let $J,\Theta $ be the dimension of the initial cell and $\Delta _{J},\Delta
_{\Theta }$ the dimension of the fix rectangular boxes. Then the length of
the basis is constant and so $vol C$ also is constant. Then if we consider
an initial rectangular box the motion will deform this cell in a
parallelogram, where the height continue to be $J$ and the base will now be $%
\Theta +\Delta \theta ,$ i. e. there is ``elongation" $\Delta \theta $ (see
figure 3), precisely%
\begin{equation*}
\Delta \theta =vt
\end{equation*}
Let us compute the evolution of $\Omega $ in this case: the number of new
boxes that appears at time $t$ will be%
\begin{equation}\label{cell1}
n=2\frac{\Delta \theta }{\Delta _{\Theta }}=2\frac{vt}{\Delta _{\Theta }}
\end{equation}
Now%
\begin{equation}\label{cell2}
\Omega =\frac{vol\Sigma }{vol C}=\frac{vol\Sigma +\Delta vol\Sigma }{volC}=%
\frac{vol\Sigma +n\Delta _{J}\Delta _{\Theta }}{volC}
\end{equation}
so%
\begin{equation}\label{cell3}
\Delta \Omega =\frac{n\Delta _{J}\Delta _{\Theta }}{volC}=\frac{n\hbar }{volC%
}=2\frac{v}{\Delta _{\Theta }}\frac{\hbar }{volC}t=2\frac{\Delta \theta }{%
\Delta _{\Theta }}\frac{\hbar }{volC}>0
\end{equation}
Then:

a.- The increment $\Delta \Omega $ is proportional to the time $t$.

b.- It is also proportional to the product of the ratio of the elongation $%
\Delta \theta $ measured in units of $\Delta _{\Theta }.$

c.- Finally it is proportional to $\frac{\hbar }{volC}$ so in the
macroscopic limit $\frac{\hbar }{volC}\rightarrow 0$ we have $\Delta \Omega
\rightarrow 0$ and the threat to CP disappears.

But the most important conclusion is that, in a generic case, even if $\frac{%
\hbar }{volC}$ would be small but if it is far from the limit $\frac{\hbar }{%
volC}\rightarrow 0,$ after enough time we will have $\Omega \gg 1$. Then the
cell ceases to be a good model for a point and it surely is the beginning of
threat to the CP. This happens even if the system is integrable, namely, $%
D_{\phi }=\Gamma $ the phase space, and the Hamiltonian $\widehat{H}=a_{0}%
\widehat{I}+a_{1}\widehat{J}+a_{2}\widehat{J}^{2}$, e. g., simply be $%
\widehat{H}=\frac{1}{2m}\widehat{P}^{2},$ namely the one of a free particle.
So fundamental graininess alone (with no chaos) can be a threat to the CP,
in the case $\frac{\hbar }{volC}>0$

\begin{figure}\label{f3}
\centering
\includegraphics{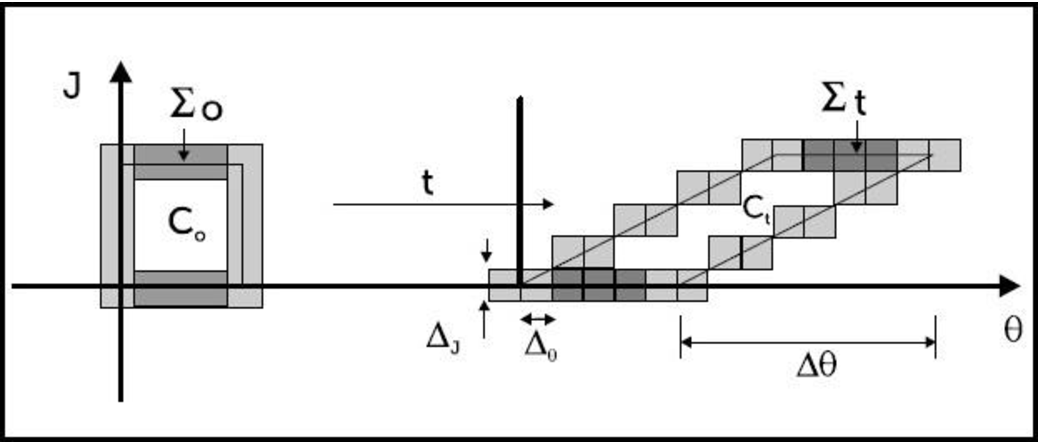}
\caption{\small{Evolution of the cell
with linear velocity.}}
\end{figure}

3.- In the most general case the Hamiltonian is $\widehat{H}=a_{0}\widehat{I}
+a_{1}\widehat{J}+a_{2}\widehat{J}^{2}+a_{3}\widehat{J}^{3}+...$ and eq. \eqref{A} becomes
\begin{equation}\label{nonlinvel}
\overline{v}(\overline{j}_{1})-\overline{v}(\overline{j_{2}})=2a_{2}(%
\overline{j}_{1}-\overline{j_{2}})+3a_{3}(\overline{j}_{1}^{2}-\overline{%
j_{2}}^{2})+...
\end{equation}
as described in figure 4 where there are not vertical deformations but there
are strong horizontal ones. Then for Hamiltonians with power bigger than 2
the threat of chaos begins.

In fact, let us consider the case
\begin{equation*}
\widehat{H}=\sum_{n=0}^{\infty }A_{n}(\widehat{J})e^{in\frac{\widehat{J}}{\Delta _{J}}}
\end{equation*}%
then
\begin{equation*}
\overline{v(j)}=\sum_{n=0}^{\infty }[A_{n}^{\prime }(\overline{j})+\frac{in}{
\Delta _{J}}A_{n}(\overline{j})]e^{in\frac{\overline{j}}{\Delta _{J}}}
=\sum_{n=0}^{\infty }B_{n}(\overline{j})e^{in\frac{\overline{j}}{\Delta
_{J}}}
\end{equation*}%
and%
\begin{equation*}
\overline{\theta }(j,t)=\overline{\theta }(j,0)+\overline{v}(\overline{j})t=%
\overline{\theta }(\overline{j},0)+t\sum_{n=0}^{\infty }B_{n}(\overline{j}%
)e^{in\frac{\overline{j}}{\Delta _{J}}}
\end{equation*}%
Then the elongation will be
\begin{equation*}
\Delta \theta =t\sum_{n=0}^{\infty }B_{n}(\overline{j})e^{in\frac{
\overline{j}}{\Delta _{J}}}
\end{equation*}
Let us consider the simple case $B_{m}(\overline{j})=const\neq 0$ and all
other $B_{n}(\overline{j})=0$ (figure 5)$,$ then
\begin{equation*}
\Delta \theta =tB_{m}e^{\frac{\overline{j}}{\Delta _{J}}}\,\,\,\,\,\, so \,\,\,\,\,\, Re(\Delta \theta) =tB_{m}\cos (m\frac{\overline{j}}{\Delta _{J}})
\end{equation*}
and the wave longitude of the oscillation of the vertical boundary curves is
$\lambda =\frac{\Delta _{J}}{m}$ and we can have $\lambda \ll \Delta _{J}$
if $m\gg 1$. Then we have%
\begin{equation*}
\Delta \Omega =\frac{\Delta vol\Sigma }{volC}=2\frac{J\Delta \theta}{J\Theta }=2\frac{B_{m}}{\Theta }t
\end{equation*}%
So when $t\rightarrow \infty $ then $\Delta \Omega \rightarrow \infty ,$ and
we have a real threat to the CP with no redemption in the classical limit.
And this can happen even in a not chaotic case since we can have $D_{\phi
}=\Gamma $ \footnote{
In if the cases 1, 2, and 3 we would take the $\widehat{H}$ as the free
variable we would have $\widehat{J}=F^{-1}(\widehat{H}),$ and, in the
corresponding figures, $H=const.$ would appear in the vertical axis, and $t,$
in the horizontal one, with the same qualitative results.}

\begin{figure}\label{f4}
\centering
\includegraphics{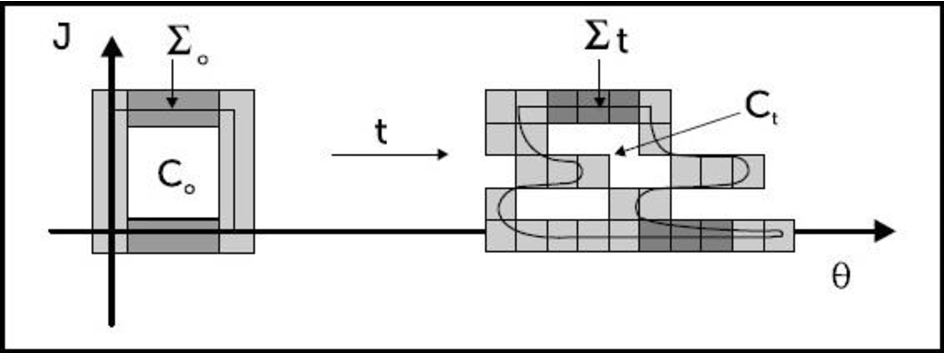}
\caption{\small{Evolution of the cell
with non linear velocity.}}
\end{figure}

4.- But things get really worst if, instead of one $D_{\phi },$ we consider
two $D_{\phi _{1}}$ and $D_{\phi ^{2}}$ and their joining zone $\mathcal{F}$%
, as in figure 6.A. Precisely let us suppose that in $D_{\phi 1}$ we have
two parallel motions and only a parallelogram deformation as in point 2, and
we use the ($\theta ,j)$ coordinate of $D_{\phi _{1}}.$ But neither in $%
\mathcal{F}$ nor in $D_{\phi _{2}}$ the just quoted coordinate $j$ is a
constant of the motion, so in $D_{\phi _{2}}$ the motion becomes completely
deformed as shown in the figure 6.A. Then if the motion goes through several
joining zones $\mathcal{F}$ it is clear that the initial regular cell will
become the amoeboid object of figure 6.B, where of course $\Omega \gg 1$.
Remember that, for the sake of simplicity, the points of all these figures
6.A and 6.B have a volume $\hbar $ (or \ really $\hbar ^{(n+1)}$ in the
general case). Then when, as a consequence of chaos, the volume of the
complex details of the amoeboid figure becomes of the order of $\hbar $ (or $%
\hbar ^{(n+1)}$ in the general case) the classical limit representing the
notion the original cell becomes meaningless as a result of chaos. Moreover
in this case we could speculate that the square box becomes strongly
deformed. But this kind of reasonings is forbidden by the Indetermination
Principle and because in our treatment square boxes are considered rigid.

Another way to see that there is a real problem is to consider that the
classical motion of the center of the initial cell (where the probabilities
to find the particle are different from zero) as the real classical motion
of a classical particle. Then in the chaotic case it may happen that at time
$t,$ the cell would get the amoeboid shape of figure 6.B. Now the center of
the original cell turns out to be outside of the amoeboid figure. Then this
center is in a zone of zero probability and cannot represent the motion of a
real point-like classical particle anymore.

So chaos and fundamental graininess are a real threat to the classical limit
of quantum mechanics and so for its interpretation.

\begin{figure}\label{f5}
\centering
\includegraphics{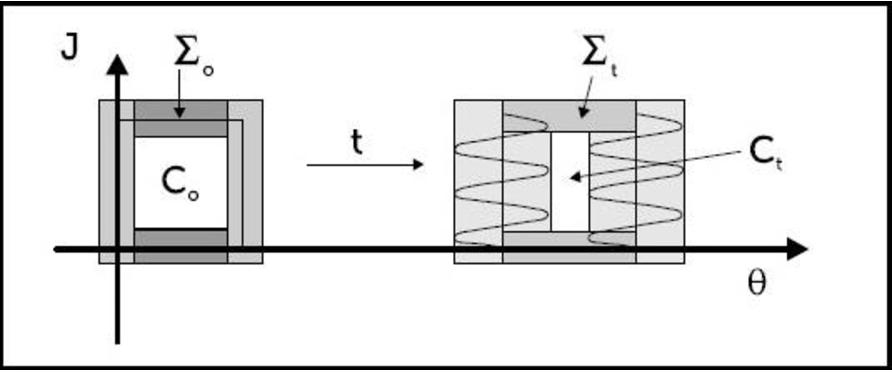}
\caption{\small{Evolution of the cell
with periodical velocity.}}
\end{figure}

\textbf{Example: the Henon-Heiles system and the high energy problem.}

In the case of Henon-Heiles classical system (\cite{Tabor} page 121) with
Hamiltonian
\begin{equation*}
H=\frac{1}{2}(p_{x}^{2}+p_{y}^{2}+x^{2}+y^{2})+x^{2}y-\frac{1}{2}y^{3}
\end{equation*}%
We can observe that:

a.-The Hamiltonian is non integrable so in the whole phase space we will
find something like figure 6.A.

b.- For energies $E=\frac{1}{12}$ (figure 44a of \cite{Tabor}) the tori are
practically unbroken, as in case 3 above. But in large $D_{\phi }$ and in a
physical case most likely $volD_{\phi }\gg \hbar ^{2}$ and CP could be far
from having practical problems with chaos at least for short periods of
time. These $D_{\phi }$ become smaller for $E=\frac{1}{8}$ (figure 44b of
\cite{Tabor}) \ and probably very tiny for $E=\frac{1}{6}$ (figure 44c of
\cite{Tabor}) so in such cases we may have serious problems with chaos (i.e.
those of case 4) since for real high energy we could have $volD_{\phi
}\approx \hbar ^{2}.$ We can obtain these conclusions because our method
allows us to evaluate the $volD_{\phi }$ on the surface defined by the
constant of motion (tori) from the Poincar\'{e} sections.

\begin{figure}\label{f6}
\centering
\includegraphics{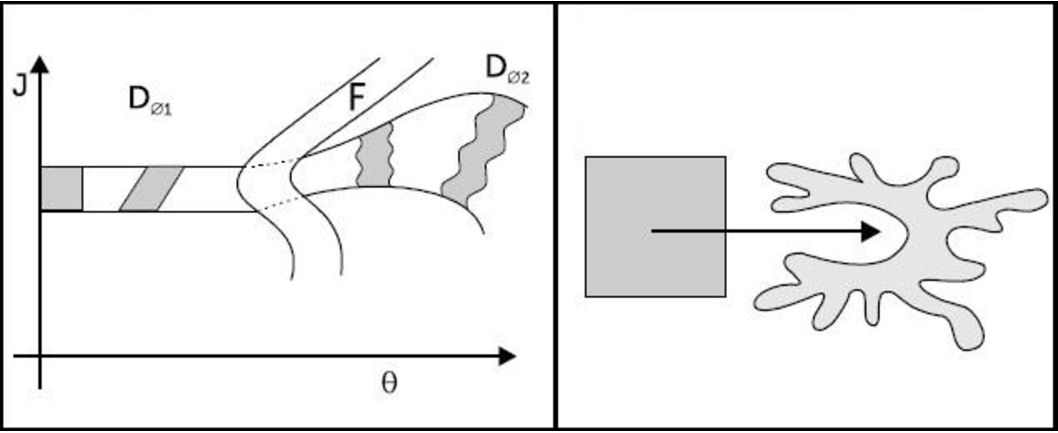}
\caption{\small{Figure 6A. A square cell scattered by a frontier. Figure 6B. A square cell becomes an ameoboidal cell.}}
\end{figure}

So we conclude that when the $D_{\phi }$ in phase space are of the order of $%
\hbar $ CP has real problems. But also we see that for high energy there is
not a generic well defined ``high energy limit". The threat of chaos to the
CP is thus explained. Moreover this example introduces the threat of chaos
to the high energy limit.
In the next section we analyze how the threat of the chaos to the CP can be suppresed taken into account the relationship between our graininess approach and the characteristic timescales of quantum chaos.

\subsection{Timescales and graininess}

As we mentioned in section 5.1 the graininess must be compatible with the quantum chaos timescales within which the typical phenomena as the statistical relaxation, the exponential localization and more generally, the instability of motion can occur. These timescales are an attempt to reconcile the discrete spectrum with the CP where the distinction between the discrete and continuous spectrum becomes relevant only for large times $t\rightarrow\infty$, see pag. 9 of \cite{casati chirikov}.

In this section, from our graininess approach we study the relations that can be obtained for the quantum chaos timescales. In section 5.3 we have seen that the condition $\Omega<1$ represents the allowed range where the notion of real point and the description of the classical trajectories become possible. The main idea is that $\Omega\ll1$ (bulky cell) implies a temporal range of validity of the fundamental graininess which can be identified with some of the characteristic timescales of quantum chaos. As in the first example of section 5.3, let us consider a two dimensional space within a domain $D_{\phi}$ and the conjugated coordinates $(J,\Theta)$ with a Hamiltonian

\begin{equation}\label{hamiltonian}
\widehat{H}=\sum_{n=0}^{\infty} a_n \widehat{J}^n
\end{equation}
In such case the difference of velocities is (see eq. \eqref{nonlinvel})

\begin{equation}\label{nonlinvel2}
v=\overline{v}(\overline{j}_{1})-\overline{v}(\overline{j_{2}})=\sum_{n=1}^{\infty} na_n \widehat{J}^{(n-1)}= 2a_{2}(%
\overline{j}_{1}-\overline{j_{2}})+3a_{3}(\overline{j}_{1}^{2}-\overline{j_{2}}^{2})+...
\end{equation}
On the other hand the evolution of $\Omega$ can be given in terms of the number of new boxes $n=n(t)$ that appear at time $t$ (see Eqs. \eqref{cell2} and \eqref{cell3})
\begin{equation}\label{nonlinvel3}
\Delta \Omega =\frac{n(t)\Delta _{J}\Delta _{\Theta }}{volC}=\frac{n(t)\hbar }{volC}
\end{equation}
We initially assume we have a bulky cell, i.e. $\Omega\ll1$. In order to obtain the characteristic timescales we only need to consider two cases: 1) Linear velocity and 2) nonlinear velocity. Let $\Omega^{\prime}$ the value of $\Omega$ at time $t$. Then by eq. \eqref{cell2} we have
\begin{equation}\label{increment1}
\Omega^{\prime}=\Omega+\Delta \Omega
\end{equation}
Since $\Omega\ll1$ if we impose that $\Omega^{\prime}=\Omega+\Delta \Omega\lesssim1$, i.e. the allowed range of the graininess, then this condition becomes into
\begin{equation}\label{increment2}
\Delta \Omega\lesssim1
\end{equation}
That is,
\begin{equation}\label{increment3}
\frac{n(t)\hbar }{volC}\lesssim1
\end{equation}
Let us see that the eq. \eqref{increment3} contains the different timescales according to the form of the Hamiltonian of the eq. \eqref{hamiltonian}. When the velocity is linear we have $a_n=0$ for all $n\geq3$ in the Hamiltonian given by the eq. \eqref{hamiltonian}. In such case we can replace eq. \eqref{cell1} in eq. \eqref{increment3} to obtain
\begin{equation}\label{increment4}
2\frac{v}{\Delta_{\Theta}}\frac{\hbar}{volC}t\lesssim1
\end{equation}
Now since $v$, $\Delta_{\Theta}$ and $volC$ are fixed, from the eq. \eqref{increment4} we have
\begin{equation}\label{increment5}
t\lesssim \left(\frac{\Delta_{\Theta}}{2v}volC\right) \hbar^{-1}=t_R \propto \hbar^{-1}
\end{equation}
Therefore we have obtained the relaxation timescale $t_R= \left(\frac{\Delta_{\Theta}}{2v}volC\right) \hbar^{-1}$ for the case of a Hamiltonian $\widehat{H}=a_0 \widehat{I}+a_1 \widehat{J}+a_2 \widehat{J}^2$ which is consistent with the so-called ``semiclassical regime" of the regular classical limits (with no chaos). In other words, for two dimensional systems our approach of the graininess plus the condition $\Delta \Omega\lesssim1$ implies a temporal range of validity of the graininess given by the relaxation timescale $t_R= \left(\frac{\Delta_{\Theta}}{2v}volC\right) \hbar^{-1}$ for the quadratic Hamiltonian case and viceversa. Moreover, from these arguments and section 5.3 it follows that there is threat to the CP only for times $t> t_R$ which are outside of the range of validity of the fundamental graininess.

Let us see what happens in the other case, i.e. when the Hamiltonian is $\widehat{H}=\sum_{n=0}^{\infty} a_n \widehat{J}^n$ with $a_n\neq0$ for some $n\geq3$. As we mentioned in the example 3 of the section 5.3 there are only strong horizontal deformations of the cells (see Fig. 4 and 5). This case includes the exponential instability where the wave-packet motion is as random as the classical trajectory and the packet is exponentially spreading with a classical rate $h$ (see pag. 14 of \cite{casati chirikov}). So we can reasonably assume\footnote{Here we are considering that the exponential spreading implies an exponential elongation of the cell as it evolves, see Fig. 4.}, hypothetically, that the number of new boxes $n=n(t)$ that appear at time $t$ is proportional to $\exp(\frac{t}{h})$ as the packet spreads, i.e.
\begin{equation}\label{increment6}
n=n(t) \propto e^{\frac{t}{h}}
\end{equation}
Then if we replace the eq. \eqref{increment6} in \eqref{increment3} we have
\begin{equation}\label{increment7}
\frac{e^{\frac{t}{h}}\hbar}{volC}\lesssim1
\end{equation}
Now applying logarithm to both sides of the eq. \eqref{increment7} we obtain
\begin{equation}\label{increment8}
\frac{t}{h}+\log(\frac{\hbar}{volC})\lesssim0
\end{equation}
That is,
\begin{equation}\label{increment9}
t\lesssim -h\log(\frac{\hbar}{volC})=\tau\propto -\log \hbar
\end{equation}
The time scale $\tau=-h\log(\frac{\hbar}{volC})$ corresponds to the logarithmic breaking time where classical and quantum mechanics agree for quantum systems with a chaotic classical behavior. In this case fundamental graininess is a real threat to CP as $t>\tau$. Given that $\tau<t_R$ then we see that the nonlinear velocity case (i.e. $a_n\neq0$ for some $n\geq3$) restricts the time range more than the linear velocity case (with no chaos). Therefore we conclude that the chaos increases the threat to the CP.

\subsection{Classical statistical limit and graininess}
We conclude with a brief discussion about the classical statistical limit of the section 4 and its relation with the non-commutative double limit of the eq. \eqref{double limit}. According to the eq. \eqref{4.4} the classical statistical limit requires the asymptotic limit $t\rightarrow\infty$ and the limit $\frac{\hbar}{S}\rightarrow0$ (see eq. \eqref{4.30}) plus the ``graininess compatibility relation" (see Eq. \eqref{increment2} or \eqref{increment3}) to guarantee that there is no threat to the CP. However, we have seen that the graininess compatibility relation leads to the different timescales of quantum chaos. Therefore, following the research line of \cite{casati chirikov} pag. 18 we should be take the two limits simultaneously but keeping the ratio $\frac{t}{t_R(q)}$ or $\frac{t}{\tau(q)}$ fixed where $q$ is the quasiclassical parameter given by $q=\frac{volC}{\hbar}$ \footnote{We assume the action $S$ proportional to $volC$ which is the volume of a given initial cell.}. From Eq. \eqref{increment5} and \eqref{increment9} we have
\begin{equation}\label{increment10}
t_R(q)=\left(\frac{\Delta_{\Theta}}{2v}\right)q^{-1}
\end{equation}
and
\begin{equation}\label{increment11}
\tau(q)=-h\log q
\end{equation}
In other words, if we take into account the graininess we must to rewrite the classical statistical limit of the eq. \eqref{4.35'} according to

\begin{equation}\label{increment12}
W-\lim_{t,q\rightarrow \infty \ , \ t\lesssim t_R(q) \ or \ \tau(q)}\rho (\phi,t)=\rho _{S}(\phi )=\rho _{\ast
}(\phi )=
\end{equation}
\begin{equation}
\sum_{i}\int_{p_{iI}}dp_{iI}^{N}\,\int_{0}^{\infty }d\omega \,\frac{%
\overline{\rho _{i}(\omega ,p_{iI})}}{C_{i}(H,P_{iI})}\,\,\,\delta (H(\phi
)-\omega )\,\delta ^{N}(P_{iI}(\phi )-p_{iI})
\end{equation}
In this manner the classical statistical limit is always compatible with the graininess and the CP is safe in all cases, regular and chaotic. On the other hand if we only take the limit $t,q\rightarrow \infty$ then we fall into the ambiguity of the non-commutative double limit given by the eq.\eqref{double limit} which, as we have seen in the sections 5.3 and 5.4, represents a threat to the CP for times that are outside of the time range of the graininess, i.e. when $t>t_R(q)$ or $t>\tau$.

\section{Conclusions}

In this paper we have:

1.- Presented a new formalism to study the classical limit of quantum
mechanics.

2.- Showed that somehow fundamental graininess alone is a threat to the CP unless the timescales of quantum chaos are taken into account (section 5.4).

3.- Demonstrated how chaos increases this threat.

4.- Proved that these threats which compromise the high energy limit of
quantum mechanics can be suppressed if we identify the bulky cell condition $\Omega\ll1$ with the quantum chaos timescales (section 5.4).

5.- Found a non trivial connection between the characteristic timescales of quantum chaos and the fundamental graininess
that allowed us to redefine a statistical classical limit that is compatible with the CP and the fundamental graininess (section 5.5).

We conclude that to avoid the threat of chaos and fundamental
graininess to the CP is necessary to take into account the characteristic timescales of quantum chaos. As we mentioned before, these timescales are an alternative solution to the ambiguity of the non commutative double limit
\begin{equation}\label{6.1}
lim_{t\rightarrow\infty} lim_{q\rightarrow\infty}\neq lim_{q\rightarrow\infty}lim_{t\rightarrow\infty}
\end{equation}
where $q$ is the quasiclassical parameter (see pag. 17 of \cite{casati chirikov}). More precisely, the mathematical need to take the limit $t\rightarrow\infty$ in the statistical classical limit (see section 4) and in asymptotic theories (e.g. ergodic theory) imply a simultaneously and conditional double limit that solves the apparent contradiction between the CP and the quantum transient pseudochaos (see pag. 18 of \cite{casati chirikov}). In our fundamental graininess approach this contradiction emerged in a geometrical way studying the domains
of definition of the constants of the motion (in the considered
non-integrable system), the corresponding broken tori at different energies
and the behavior of the cells for different Hamiltonians (as in case 1,2,
and 3 of section 5.3). In section 5.5 considering an initial bulky cell $\Omega\ll1$, the compatibility condition $\Delta \Omega\lesssim1$ and taking into account the statistical classical limit of section 4 we translated these finite time intervals of ``quantum pseudo chaos" to a classical limit that is compatible with the CP and the general structure of classically chaotic quantum motion (see Fig. 5 of \cite{casati chirikov}).
In this sense we conclude that the fundamental graininess plus the statistical classical limit provide a new formalism to study the classical limit that is compatible with the CP and the quantum chaos timescales. In the next table we summarize these results.

\vskip0.2truecm
\centerline{TABLE I: Fundamental graininess, statistical classical limit, and their relationships\footnote{By ``undefined" we mean the absence of this element within the formalism.}}\vskip0.2truecm

\begin{tabular}{||l |c| r||}
\hline
\hline
   &            &            \\
   &  & \,\,\,\,\,\, Fundamental graininess \,\,\,\,\,\,\,\,\,  \\
             Statistical classical limit (only)  & Fundamental graininess (only) &     + \,\,\,\,\,\,\,\,\,\,\,\,\,\,\,\,\,\,\,\,\,\,\,\,\,\,\,\,\,\,\,\,\,\,\,\,\,\      \\
              &            &     \,\,\,\,\,\, Statistical classical limit  \,\,\,\,\,\,     \\
\hline
            &            &            \\
    \,\  $lim_{q\rightarrow\infty} lim_{t\rightarrow\infty}$      &    &    \,\ $\lim_{t,q\rightarrow \infty \ , \ t\lesssim t_R(q) \ or \ \tau(q)}$     \,\,\,\,\,\,\,\,\,  \\
       \,\,\,\,\,\,(quantum behavior    &  \,\ undefined classical limit &     (double limit taken  \,\,\,\,\,\,\,\,\,\,\,\,\,\,\,\,\,\,\,\,\,\,\,\,\,\,\,     \\
      \,\,\,\,\,\, with no chaos at all) &            &     simultaneously, chaotic \,\,\,\,\,\,\,\,\,\,\,\,\,\,\,\,\,\,       \\
  && quantum motion) \,\,\,\,\,\,\,\,\,\,\,\,\,\,\,\,\,\,\\

           \hline
           &            &            \\
      \,\ infinite relaxation time    &   \,\ finite relaxation time \,\,\,\,\,\,\,\,\,\,\,\, &    \,\ finite relaxation time \,\,\,\,\,\,\,\,\,\,\,\,\,\,\,\,\,      \\
           \,\,\,\,\,\, $t_R=\infty$     &    $t_R(q)=\left(\frac{\Delta_{\Theta}}{2v}\right)q^{-1}$  \,\,\,\,\,\,\,\,\,\,\,\,       &   $t_R(q)=\left(\frac{\Delta_{\Theta}}{2v}\right)q^{-1}$  \,\,\,\,\,\,\,\,\,\,\,\,\,\,\,\,\,\,\,\,\,\,\,\,       \\
           &(two dimensional phase space)&(two dimensional phase space)\\
              \hline
           &            &            \\
         \,\,\,\,\,\,\,\,\,\,\ undefined timescales   &    \,\ defined quantum chaos \,\,\,\,\,\,  &      \,\ defined quantum chaos \,\,\,\,\,\,\,\,\,\,\,    \\
              &      timescales $t_R(q)$ and $\tau(q)$   &    timescales $t_R(q)$ and $\tau(q)$  \,\,\,\,\,      \\
              &(two dimensional phase space)&(two dimensional phase space)\\
              \hline
           &            &            \\
        \,\,\,\,\,\,\,\,\,\,\,\,\,\,\ threat to the CP    &  no threat to the CP &   no threat to the CP \,\,\,\,\,\,\,\,\,\,\,\,\,\,\ \\
              &            &            \\
              \hline
           &            &            \\
       \,\ $\rho_{\ast}(\phi)=lim_{q\rightarrow\infty} lim_{t\rightarrow\infty}\rho(\phi,t)$       &     &   $\rho_{\ast}(\phi)=$ \,\,\,\,\,\,\,\,\,\,\,\,\,\,\,\,\,\,\,\,\,\,\,\,\,\,\,\,\,\,\,\,\,\,\,\,\,\,\,\,\,\,\,\,\,\,\,\,\,\,\,\,\,\,\,\,\,\,  \\
       && \\
          non compatible weak limit with    & undefined weak limit   &  $\lim_{t,q\rightarrow \infty \ , \ t\lesssim t_R(q) \ or \ \tau(q)}\rho(\phi,t)$\\
              &        &    \\
           \,\,\,\,\,\,\,\,\,\,\,\,\,\,\,\,\,\,\,\,\,\,\,\,\, the CP   &        &   compatible weak limit with the   \\
              &        &     fundamental graininess and the  \\
              &            &       CP   \,\,\,\,\,\,\,\,\,\,\,\,\,\,\,\,\,\,\,\,\,\,\,\,\,\,\,\,\,\,\,\,\,\,\  \\
\hline
\hline
\end{tabular}

\vspace{0.5cm}
\noindent From the Table I we can see how the fundamental graininess and the statistical classical limit complement their indefinite sectors (rows) to give rise to a better classical limit that is compatible with the fundamental graininess and where the CP is safe (third column).
Also, it should be noted that at least for the quadratic Hamiltonian case $\widehat{H}=a_0 \widehat{I}+a_1 \widehat{J}+a_2 \widehat{J}^2$ (linear velocity, see example 2 of section 5.3), the results of the Table I can be generalized for an phase space of any finite dimension. Consider that the dimension of the phase space is $2D$. Let $\Delta J_1,...,\Delta J_D,\Delta \Theta_1,...,\Delta \Theta_D$ be the size of the fix rectangular boxes. Then following the arguments of the example 2 of section 5.3 we have an ``elongation" at time $t$

\begin{equation}\label{6.2}
\Delta \theta_1 \Delta \theta_2...\Delta \theta_D=vt
\end{equation}
where $v=2a_2(\overline{j}_1-\overline{j}_2)=\overline{v}(\overline{j}_1)-\overline{v}(\overline{j}_2)$, see eq. \eqref{B}. In this case we have ``elongations" in each of the $D$ directions, i.e. for each direction $j$ with $j=1,...,D$ we have a stretch like the Fig. 3. Therefore, the increment $\Delta \Omega$ at time $t$ will be\footnote{Here we use that $\Delta J_1...\Delta J_D\Delta\Theta_1...\Delta \Theta_D\thicksim\hbar^{D}$ which represents a small square in a 2D-dimensional phase space.}
\begin{equation}\label{6.3}
\Delta \Omega=n\frac{\Delta J_1...\Delta J_D\Delta\Theta_1...\Delta \Theta_D}{vol C}=n\frac{\hbar^{D}}{vol C}
\end{equation}

\noindent where $n=n(D)$ is the number of new boxes that appears at time $t$ which depends on the dimension $D$ of the phase space. Moreover, as in the example 2 of section 5.3. since the velocity is linear then $n$ is proportional to the time $t$. Then we have
\begin{equation}\label{6.4}
n=\alpha(D,\Delta J_1,...,\Delta J_D,\Delta \Theta_1,...,\Delta \Theta_D)t \,\,\,\,\,\,\,\, for \,\ some \,\ \alpha\in\mathbb{R}
\end{equation}
\noindent Now, by replacing the eq. \eqref{6.4} in the eq. \eqref{6.3} and assuming the graininess condition $\Delta \Omega\lesssim1$ (see eq. \eqref{increment2}) we obtain
\begin{equation}\label{6.5}
\Delta \Omega=\alpha t\frac{\hbar^{D}}{vol C}\lesssim1\,\,\,\, \Longrightarrow \,\,\,\, t\lesssim \left(\alpha^{-1}volC \right)\hbar^{-D}=t_R\propto \hbar^{-D}
\end{equation}
\noindent which is the relaxation timescale $t_R$ for the case of a quadratic Hamiltonian with a phase space of dimension 2D.

\bigskip Finally, taking into account the fundamental graininess and based in these results we could go on with the following
speculation: In the classical level, the KAM theorem was the solution of the
problem of the scarcity of chaos in the solar system, since the tori were
broken but not badly broken. In the same way we could consider that the
study of the size of the $D_{\phi _{i}}$, for different levels of energy,
could also explain the behavior of chaotic quantum systems and may be the
scarcity of chaos in these systems. I. e. it may be that, many cases, the $%
D_{\phi _{i}}$ would be large enough to endow these systems with a
quasi-integral chaotic behavior Along these lines we will continue our
research.

\textbf{Acknowledgements: }This work was partially supported by grants of
the Buenos Aires University, the CONICET (Argentine Research Council) and
FONCYT (Argentine Found for Science and Technology).

\end{document}